\documentclass[prd,showpacs,amsmath,amssymb,nofootinbib, 12pt]{revtex4}
\usepackage{epsf,epsfig}

\def\be{\begin{eqnarray}}
\def\en{\end{eqnarray}}
\def\non{\nonumber}
\def\nslash{\rlap{\hspace{0.01cm}/}{n}}
\def\vslash{\rlap{\hspace{-0.02cm}/}{v}}
\def\pslash{\rlap{\hspace{0.03cm}/}{p}}
\def\dslash{\rlap{\hspace{0.08cm}/}{D}}
\def\hl{{heavy-to-light~}}

\def\lc{{light cone~}}

\def\lqcd{\Lambda_{\rm QCD}}
\def\la{\langle}
\def\ra{\rangle}

\def\npb{{ Nucl. Phys. B}~}

\begin{document}

\title{Heavy-to-light form factors on the light cone}

\author{ Cai-Dian L\"u$^a$\footnote{E-mail: lucd@mail.ihep.ac.cn},
         Wei Wang$^a$\footnote{E-mail: wwang@mail.ihep.ac.cn} and
         Zheng-Tao Wei$^{b,c,d}$\footnote{E-mail: weizt@nankai.edu.cn}  }
\affiliation{{\it \small $^a$  Institute of High Energy
Physics, CAS, P.O.Box 918(4), Beijing 100049, China} \\
{\it \small$^b$  Department of Physics, Nankai University, Tianjin 300071, China} \\
{\it \small $^c$  National Center for Theoretical Sciences, National Cheng-Kung
                  University, Tainan 701, Taiwan} \\
{\it \small $^d$  Institute of Physics, Academia Sinica, Taipei
                  115, Taiwan}}

\begin{abstract}

The light cone method provides a convenient non-perturbative tool to study the heavy-to-light
form factors. We construct a light cone quark model utilizing the soft collinear effective
theory. In the leading order of effective theory, the form factors for $B$ to light
pseudoscalar and vector mesons are reduced to three universal form factors which can be
calculated as overlaps of hadron light cone wave functions. The numerical results show that
the leading contribution is close to the results from other approaches. The $q^2$ dependence
of the heavy-to-light form factors are also presented.
\end{abstract}


\maketitle

\newpage

\section{Introduction}\label{intro}

The hadronic matrix elements of weak $B$ decays to a light pseudoscalar (P) and to a vector
meson (V) are described by $B\to P$ and $B\to V$ transition form factors, respectively. These
heavy-to-light form factors are essential to study the semileptonic and even non-leptonic $B$
decays. Information on the form factors is crucial to test the mechanism of CP violation in
the Standard Model and to extract the CKM parameters \cite{CKM}. For instance, the
$B\to\pi(\rho)$ form factors are required to determine the CKM matrix element $|V_{ub}|$
precisely. In $B\to V\gamma$ and $B\to Vl^+l^-$ processes which are sensitive to new physics,
the precise evaluation of $B\to V$ form factors is indispensable. Another interesting reason
for the study of the \hl form factors is that they provide an ideal laboratory to explore the
rich structures of QCD dynamics. At the large recoil region where the final state light meson
moves fast, the \hl system contains internal information on both short and long distance QCD
dynamics with the factorization theorem.

There are already many methods calculating the heavy-to-light transition form factors in the
literature such as simple quark model \cite{bsw}, the light cone quark models (LCQM)
\cite{Jaus1,CCH1,CJK,CCH2} \footnote{In some references, the authors prefer to use the term
``light front". We will use the term of ``light cone" which is widely adopted in SCET, LCSR
and other approaches.}, the light cone sum rules (LCSR) \cite{LCSRP,BZpseudo,BZvector}, the
perturbative QCD (PQCD) approach based on $k_T$ factorization \cite{PQCD} etc.

In Ref.~\cite{CYOPR}, a model-independent way to look for relations between different form
factors is suggested by analogy with the heavy-to-heavy transitions \cite{IW}. One important
observation is that in the heavy quark mass and large energy of light meson limit, the spin
symmetry relates the form factors for $B\to P$ and $B\to V$ to three universal
energy-dependent functions: $\zeta_P$ for pseudoscalar meson; and $\zeta_{||},~\zeta_{\bot}$
for longitudinally and transversely polarized vector meson, respectively. The development of
soft collinear effective theory (SCET) makes the analysis on a more rigorous foundation.  The
SCET is a powerful method to systematically separate the dynamics at different scales: hard
scale $m_b$ ($b$ quark mass), hard intermediate scale $\mu_{hc}=\sqrt{m_b\lqcd}$, soft scale
$\lqcd$ and to sum large logs using the renormalization group technics. After a series of
researches \cite{SCETff1,SCETff2,BCDF,SCETff3,SCETff4}, a factorization formula is established
for the \hl form factors in the heavy quark mass and large energy limit as
 \be
 F_i(q^2)=C_i(E,\mu_I)\zeta_j(\mu_I,E)+\phi_B(\omega,\mu_{II})
 \otimes T_i(E,u,\omega,\mu_{II}) \otimes \phi_M(u,\mu_{II}),
 \en
where the indices $j$ represent ${(P,||,\bot)}$ and $\otimes$ denotes the convolutions over
light cone momentum factions. $\phi_B(\omega)$ and $\phi_M(u)$ are light cone distribution
amplitudes for $\bar B$ and light mesons. The coefficients $C_i$ and $T_i$ are perturbatively
calculable functions which include hard gluon corrections. The functions $\zeta_j$ denote the
universal functions that satisfies the spin symmetry.

Although soft collinear effective theory is really powerful and rigorous, the form factors
$\zeta_j$ cannot be directly calculated. These functions are non-perturbative in principle and
the evaluation of them relies on non-perturbative methods. Lattice simulation on
heavy-to-light form factors is usually restricted to the region with final meson energy
$E<1\mbox{ GeV}$ and cannot be applied to our case directly where the light meson carries the
energy of order $M_B/2$\footnote{ This may be changed by applying ``moving" NRQCD in lattice
QCD \cite{movingNRQCD}. For a recent development, please see \cite{movingDWL} and references
therein.}. The construction of LCSR within SCET has been explored recently in
\cite{SCETLCSR1,SCETLCSR2,SCETLCSR3}. In these studies, only the pseudoscalar meson form
factor $\zeta_P$ are calculated at present.

The light cone field theory provides another natural language to study these processes. As
pointed out in \cite{BPP}, light cone QCD has some unique features which are particularly
suitable to describe a hadronic bound state. For instance, the vacuum state in this approach
is much simpler than that in other approaches. The light cone wave functions, which describe
the hadron in terms of their fundamental quark and gluon degrees of freedom, are independent
of the hadron momentum and thus are explicitly Lorentz invariant. The light cone Fock space
expansion provides a complete relativistic many-particle basis for a hadron. For hard
exclusive processes with large momentum transfer, factorization theorem in the perturbative
light cone QCD makes first-principle predictions \cite{LB}. For non-perturbative QCD,  an
approach which combines the advantage of light cone method with the low energy constituent
quark model is more appealing. This approach, which we will call light cone quark model
(LCQM), has been successfully applied to the calculation of the meson decay constants and
hadronic form factors \cite{Jaus1,Jaus2,CCH1,CJK,CCH2,LFHW}.

As far as the form factors are concerned, they can be generally represented by the convolution
of $B$ and light meson wave functions in the light cone approach as
 \be
  F(q^2)=\sum\limits_{n_1, n_2}\int\left\{\overline{\prod \limits_i}
   \frac{dx_id^2k_{\bot i}}{16\pi^3}\right\}
   \left\{\overline{\prod\limits_{j}}\frac{dx_jd^2k_{\bot j}}{16\pi^3}\right\}
   \Psi_M^{(n_1)*}(x_i,k_{\bot i})\Psi_B^{(n_2)}(x_j,k_{\bot j}).
 \en
where the sum is over all Fock states with $n_1,~n_2$ the particle numbers, $i,j$ denote the
$i$- and $j$-th constitutes of the light meson and $\bar B$ meson, respectively. The product
is performed over the longitudinal momentum fractions $x_{i,j}$ and the transverse momenta
$k_{\bot i,j}$. The light cone wave function $\Psi(x,k_{\bot})$ is the generalization of
distribution amplitude $\phi(x)$ by including the transverse momentum distributions. This
formulation contains both hard and soft interactions.

The main purpose of this paper is to develop a non-perturbative light cone approach within the
soft collinear effective theory and to evaluate the three universal \hl form factors directly.
The close relation between the light cone QCD and soft collinear effective theory was noted in
\cite{BCDF}. The SCET has the advantage that a systematic power expansion with small parameter
$\lqcd/m_b$ (or $\sqrt{\lqcd/m_b}$) can be performed to improve the calculation accuracy order
by order. The combination of the two methods can reduce the model dependence of
non-perturbative methods. In the conventional light cone approach, all the quarks are
on-shell. Now in the new approach, it is convenient to choose the light energetic quark as the
collinear mode in the soft collinear effective theory and the heavy quark field as that in the
heavy quark effective theory. The spectator antiquark is remained as the soft mode. By this
way, the light cone quark model within the soft collinear effective theory is established.
Then we can calculate the $B\to P$ and $B\to V$  form factors  order by order.

The paper is organized as follows. In Section~\ref{section2}, we first present the definition
of the three universal form factors from the spin symmetry relations. We then discuss a light
cone quark model within soft collinear effective theory. The numerical results for the form
factors and discussions are presented in Section~\ref{section3}. The final part contains our
conclusion.


\section{The  heavy-to-light form factors in the light cone
approach}\label{section2}

\subsection{Definitions of the \hl   form factors}

The $\bar B\to P$ and $\bar B\to V$ form factors are defined under
the conventional form as follows
 \begin{eqnarray}
  \la P(P')|\bar q\gamma^{\mu}b|\bar B(P)\ra
   &=&f_+(q^2)\left[ (P+P')^{\mu}-\frac{M_B^2-M_P^2}{q^2}q^{\mu} \right]
     +f_0(q^2)\frac{M_B^2-M_P^2}{q^2}q^{\mu}, \non \\
  \la P(P')|\bar q\sigma^{\mu\nu}q_{\nu}b|\bar B(P)\ra
   &=&i\frac{f_T(q^2)}{M_B+M_P)}
    \left[ q^2(P+P')^{\mu}-(M_B^2-M_P^2)q^{\mu} \right], \non \\
  \la V(P',\epsilon^*)|\bar q\gamma^{\mu}b|\bar B(P)\ra
   &=&-\frac{2V(q^2)}{M_B+M_V}\epsilon^{\mu\nu\rho\sigma}
     \epsilon^*_{\nu}P_{\rho}P'_{\sigma}, \non\\
  \la V(P',\epsilon^*)|\bar q\gamma^{\mu}\gamma_5 b|\bar B(P)\ra
   &=&2iM_V A_0(q^2)\frac{\epsilon^*\cdot q}{q^2}q^{\mu}
    +i(M_B+M_V)A_1(q^2)\left[\epsilon^*_{\mu}
    -\frac{\epsilon^*\cdot q}{q^2}q^{\mu} \right] \non\\
    &&-iA_2(q^2)\frac{\epsilon^*\cdot q}{M_B+M_V}
     \left[ (P+P')^{\mu}-\frac{M_B^2-M_V^2}{q^2}q^{\mu} \right],\non\\
  \la V(P',\epsilon^*)|\bar q\sigma^{\mu\nu}q_{\nu}b|\bar B(P)\ra
   &=&-2iT_1(q^2)\epsilon^{\mu\nu\rho\sigma}
     \epsilon^*_{\nu}P_{\rho}P'_{\sigma}, \non\\
  \la V(P',\epsilon^*)|\bar q\sigma^{\mu\nu}\gamma_5q_{\nu}b|\bar B(P)\ra
   &=&iT_2(q^2)\left[(M_B^2-M_V^2)\epsilon^{*\mu}
       -(\epsilon^*\cdot q)(P+P')^{\mu} \right]\non\\
   &&+iT_3(q^2)(\epsilon^*\cdot q)\left[
       q^{\mu}-\frac{q^2}{M_B^2-M_V^2}(P+P')^{\mu}\right],
 \end{eqnarray}
where $q=P-P'$ is the momentum transfer, $M_B$ the $\bar B$ meson mass, $M_{P,V}$ the mass of
the pseudoscalar and vector mesons, $\epsilon$ the polarization vector of the vector meson. We
have used the convention $\epsilon^{0123}=+1$. In the following, we choose the convention
within which the vectors $n_{\pm}$ are $n_+^{\mu}=(1,0,0,-1)$, $n_-^{\mu}=(1,0,0,1)$ and the
light cone momentum components are $p^+=n_+\cdot p=p^0+p^3$, $p^-=n_-\cdot p=p^0-p^3$,
$p_b=m_b v$. Our convention for the vectors $n_{\pm}$ are different from that in most
literatures. In the above definitions, there are ten form factors in total: $f_+$, $f_0$,
$f_T$ for the pseudoscalar meson; $V,~A_0,~A_1,~A_2,~,T_1,~T_2,~T_3$ for the vector meson.
Note that the form factors are in general different for each hadron.

In SCET, the energetic light quark is described by its leading two component spinor
$\xi=\frac{\nslash_-\nslash_+}{4}q$ and the heavy quark is replaced by $h_v=e^{im_bv\cdot
x}\frac{(1+\vslash)}{2}b$. The weak current $\bar q\Gamma b$ in full QCD is matched onto the
SCET current $\bar\xi \Gamma h_v$ at tree level where we have omitted the Wilson lines for
simplicity. For an arbitrary matrix $\Gamma$, $\bar\xi \Gamma h_v$ has only three independent
Dirac structures. One convenient choice is discussed in Ref.~\cite{SCETff1,relation}: $\bar\xi
h_v$, $\bar\xi \gamma_5 h_v$ and $\bar\xi \gamma_{\bot}^{\mu}h_v$. It can be seen from a trace
technology by
 \be
  \frac{\nslash_+\nslash_-}{4} \Gamma \frac{(1+\vslash)}{2}=
  \frac{\nslash_+\nslash_-}{4}\left[c_1+c_2\gamma_5+
   c_3\gamma_{\bot}^{\mu}\right]\frac{(1+\vslash)}{2},
 \en
where $c_i$s are defined as:
 \be
  &&c_1=\frac{1}{4}{\rm Tr}\left[(1+\vslash)\nslash_-\Gamma\right],
  \qquad \qquad
  c_2=\frac{1}{4}{\rm Tr}\left[(1+\vslash)\nslash_-\gamma_5
   \Gamma\right], \non\\
  &&c_3=\frac{1}{4}{\rm Tr}\left[(1+\vslash)\nslash_-
   \gamma_{\bot\mu}\Gamma\right].
 \en

The above spin symmetry leads to non-trivial relations for the \hl form factors: the ten form
factors are reduced to three independent universal form factors.  The $B$ to light universal
form factors $\zeta_P$, $\zeta_{||,\perp}$ are defined as
 \be \label{eq:softFF}
  \la P|\bar \xi h_v|\bar B\ra &=& 2E\zeta_P(E), \non\\
  \la V|\bar \xi\gamma_5 h_v|\bar B\ra &=&
   -2iM_V\zeta_{||}(E)v\cdot \epsilon^*, \non\\
  \la V|\xi\gamma^\mu_\perp h_v|\bar B\ra &=&
   -2E\zeta_\perp(E)\epsilon^{\mu\nu\rho\sigma}
   \epsilon_\nu^* v_\rho n_{-\sigma},
 \en
where $E=(M_B^2-q^2)/2M_B$ is the energy of the light meson (neglecting the small mass of the
final state meson) and $q$ is the momentum transfer. $\zeta_{i(i=P,||,\zeta)}$ are functions
of energy of the light meson. Up to leading order of $\alpha_s$ and leading power of
$\lqcd/m_b$, the total ten physical form factors are determined from the three independent
factors to the leading order of $\alpha_s$ as
 \be
 &&f_+(q^2)=\frac{M_B}{2E}f_0(q^2)=
   \frac{M_B}{M_B+M_P}f_T(q^2)=\zeta_P(E),\non\\
 &&\frac{M_B}{M_B+M_V}V(q^2)=
   \frac{M_B+M_V}{2E}A_1(q^2)=\zeta_{\bot}(E), \non\\
 &&A_0(q^2)=\zeta_{||}(E), \qquad
   A_2(q^2)=\frac{M_B}{M_B-M_V}
   \left[\zeta_{\bot}(E)-\frac{M_V}{E}\zeta_{||}(E)\right], \non\\
 &&T_1(q^2)=\frac{M_B}{2E}T_2(q^2)=\zeta_{\bot}(E), \qquad
   T_3(q^2)=\zeta_{\bot}(E)-\frac{M_V}{E}\zeta_{||}(E).\label{reduction}
 \en
As in \cite{CYOPR, BF1}, we keep the leading kinematic light meson
mass correction and neglect the higher $M_{P,V}^2/M_B^2$ terms.

\subsection{Light cone quark model}

We start with a discussion of hadron bound states on the light cone. The goal is to find a
relativistic invariant description of the hadron in terms of its fundamental quark and gluon
constitutes. For a complete Fock state basis $|n\ra$, the hadron is expanded by a series of
wave functions: $|h\ra=\sum\limits_n |n\ra\la n|h\ra=\sum\limits_n |n\ra\psi_{n/h}$. It is
convenient to use a light cone Fock state basis on which the hadron with momentum $\tilde
P=(P^+, P_{\bot})$ is described by \cite{BPP}
 \be
 |h:\tilde P\ra=\sum\limits_{n,\lambda_i}\int\left\{\overline{\prod \limits_i}
   \frac{dx_id^2k_{\bot i}}{\sqrt{x_i}16\pi^3}\right\}|n:x_iP^+,x_iP_{\bot
  i}+k_{\bot i},\lambda_i\ra \Psi_{n/h}(x_i,k_{\bot i},\lambda_i),
 \en
where the sum is overl all Fock states and helicities and the product is performed on the
variables $x_i$ and $k_{\bot i}$ not on the wave functions $\Psi_{n/h}(x_i,k_{\bot
i},\lambda_i)$,
 \be
 \overline{\prod \limits_i}dx_id^2k_{\bot i}=\prod\limits_i dx_id^2k_{\bot i}
 \delta\left(1-\sum\limits_j x_j\right)16\pi^3\delta^2\left(\sum\limits_j k_{\bot j}\right).
 \en
The essential variables are boost-invariant light cone momentum fractions $x_i=p_i^+/P^+$ with
$p_i$ momenta of quark or gluon and the internal transverse momenta $k_{\bot i}=p_{\bot i}-x_i
P_{\bot}$. The light cone momentum fractions $x_i$ and the internal transverse momenta
$k_{\bot i}$ are relative variables which are independent of the hadron momentum. The wave
functions in terms of these variables are explicitly Lorentz invariant and they are the
probability amplitudes for finding $n$ partons with momentum fractions $x_i$ and relative
momentum $k_{\bot i}$ in the hadron. The total probability equals to 1 which implies a
normalization condition
 \be
  \sum\limits_{n,\lambda_i}\int\left\{\overline{\prod \limits_i}
   \frac{dx_id^2k_{\bot i}}{16\pi^3}\right\}
   |\Psi_{n/h}(x_i,k_{\bot i},\lambda_i)|^2=1.
 \en

The hadron state $|h\ra$ is the eigenstate of light cone Hamiltonian $H_{LC}|h \ra=M^2|h\ra$
with  the hadron mass $M$. Solving the eigenstate equation with the full Fock states is very
difficult which is beyond our capability. We will meet an infinite number of coupled equations
and the problems of some nonphysical singularities (endpoint singularities $x_i\to 0$ or
ultraviolet singularities $k_{\bot}\to \infty$). What concerns us most is the wave function at
the endpoint region. For the wave functions $\Psi_{n/h}(x_i,k_{\bot i},\lambda_i)$, one
general property is found \cite{BPP}
 \be
 \Psi_{n/h}(x_i,k_{\bot i},\lambda_i)\to 0~~~~~{\rm as}~x_i\to 0.
 \en
This constraint means that the probability of finding partons with very small longitudinal
momentum is little. In this mechanism, the $\bar B$ meson wave function is overlapped with the
light meson wave function at the endpoint where the valence antiquark carries momentum of
order of the hadron scale. In the infinite heavy quark mass limit, the light meson wave
functions at the endpoint are suppressed. However, at the realistic $m_b$ scale, the
suppression is not so heavy, that soft contribution still dominates the \hl form factors.

The solution of all wave functions from first principle is not obtainable at present. We will
use the constituent quark model. The constituent quark masses are about several hundred MeV
for light quarks which are much larger than the current quark mass obtained from the chiral
perturbation theory. The appreciable mass absorbs dynamical effects from complicated vacuum in
the common instanton form \cite{WWHZP}. A key approximation adopted in the light cone quark
model is the mock-hadron approximation \cite{HI} where the hadron is dominated by the lowest
Fock state with free quarks. Under the valence quark assumption, we can write a meson state
$M$ constituting a quark $q_1$ and an antiquark $\bar q_2$ by
 \be
 |M(P,S,S_z)\ra=\int\frac{dp_1^+d^2p_{1\bot}}{16\pi^3}
   \frac{dp_2^+d^2p_{2\bot}}{16\pi^3}16\pi^3
   \delta^3(\tilde P-\tilde p_1-\tilde p_2) \non\\
   \times \sum\limits_{\lambda_1,\lambda_2}\Psi^{SS_z}
   (\tilde p_1,\tilde p_2,\lambda_1,\lambda_2)
   |q_1(p_1,\lambda)\bar q_2(p_2,\lambda)\ra,
 \en
where the meson denoted by its   momentum $P$ and spin $S$, $S_z$, the constituent quarks
$q_1(\bar q_2)$ denoted by momenta $p_1(p_2)$ and the light cone helicities
$\lambda_1(\lambda_2)$. The 4-momentum $p$ is defined as
 \be
  \tilde p=(p^+,~p_\perp), \qquad p_\perp=(p^1,~p^2),\qquad
  p^-=\frac{m^2+p_\perp^2}{p^+}.
 \en
From the momentum, we can see that the quarks in the meson are taken to be on the mass shell.
In the following, we choose a frame where the transverse momentum of the meson is zero, {\it
i.e.}, $P_\perp=0$. The light-front momenta $p_1$ and $p_2$ in terms of light cone variables
are
 \be
  p_1^+=x_1P^+, \qquad p_2^+=x_2P^+, \qquad
  p_{1\perp}=-p_{2\perp}=k_\perp,
  \en
where $x_i$ are the light cone momentum fractions and they satisfy $0<x_1, x_2<1$ and
$x_1+x_2=1$. The invariant mass $M_0=p_1+p_2$ of the constituents and the relative momentum
$p_z$ in $z$ direction can be written as
 \be
  M^2_0=\frac{m_1^2+k_\perp^2}{x_1}+\frac{m_2^2+k_\perp^2}{x_2},\qquad
  p_z=\frac{x_2 M_0}{2}-\frac{m_2^2+k_{\bot}^2}{2x_2M_0}.
 \en
Note that the invariant mass of the quark system is different from the meson total momentum,
i.e. $p_1+p_2\neq P$.

The momentum-space wave function related to the meson bound state can be expressed as
 \be \Psi^{SS_z} (p_1,p_2,\lambda_1,\lambda_2)=
  R^{SS_z}_{\lambda_1\lambda_2}(x,k_\perp)\phi(x,k_\perp),
 \en
where the $\phi(x,k_\perp)$ describes the momentum distribution of the constituents in the
bound state with $x\equiv x_2$, and $R^{SS_z}_{\lambda_1,\lambda_2}$ constructs a state of
definite spin $(S,S_z)$ out of the light cone helicity $(\lambda_1,\lambda_2)$ eigenstates. In
practice, it is convenient to use the covariant form for $R^{SS_z}_{\lambda_1,\lambda_2}$
\cite{Jaus1,Jaus2}:
 \be
  R^{SS_z}_{\lambda_1,\lambda_2}(x,k_\perp)=
   \frac{\sqrt{p_1^+p_2^+}}{\sqrt 2 \tilde M_0}
   \bar u(p_1,\lambda_1)\Gamma v(p_2,\lambda_2),
 \en
where the parameter $\tilde M_0\equiv \sqrt{M_0^2-(m_1-m_2)^2}$ and the $\Gamma$ matrices for
the mesons are defined as
 \be
 \Gamma_P&=&-i\frac{\gamma_5}{\sqrt N_c},\qquad\qquad\qquad~~~~
  {\rm for~ pseudoscalar~meson}\non ,\\
 \Gamma_V&=&\frac{-\hat\epsilon\!\!\!\slash(S_z)+\frac{\hat
  \epsilon\cdot (p_1-p_2)}{M_0+m_1+m_2}}{\sqrt N_c},~~~~~
   {\rm for~ vector~ meson}
 \en
with $N_c=3$.   The transverse and longitudinal polarization vectors $\hat \epsilon$ are:
 \be
  \hat\epsilon^\mu(\pm1)=\Big(0,~0,~
   \vec\epsilon_\perp(\pm1) \Big), \qquad
  \hat \epsilon^\mu(0)=\frac{1}{M_0}\Big(
   -\frac{M_0^2}{P^+},~P^+,~0\Big),
 \en
where $\vec\epsilon_\perp(\pm1)=\mp(1,\pm i)/\sqrt 2$. The Dirac spinors satisfy the relation:
 \be \label{eq:spin}
 &&\sum\limits_{\lambda}u(p,\lambda)\bar u(p,\lambda)=\frac{
  (\pslash+m)}{p^+}, \qquad {\rm for ~ quark} ,\non\\
 &&\sum\limits_{\lambda}v(p,\lambda)\bar v(p,\lambda)=
  \frac{(\pslash-m)}{p^+}, \qquad {\rm for ~ antiquark} .
 \en
The momentum distribution amplitude $\phi(x, k_\bot)$ is the generalization of the
distribution amplitude $\phi(x)$ which is normalized as
 \be \label{eq:Norm1}
 \int \frac{dx d^2 k_{\bot}}{2 (2\pi)^3}
   |\phi(x,k_{\bot})|^2=1.
 \en

Before discussing of the form factors, we will study the decay constants in the light cone
approach. The decay constants $f_{P,V}$ are defined by the matrix elements of the axial-vector
current for pseudoscalar meson and the vector current for vector meson:
 \be
 \la 0|A^\mu|P(P)\ra=if_PP^\mu, \qquad
 \la 0|V^\mu|V(P)\ra=M_Vf_V\epsilon^\mu,
 \en
where $P$ is the meson momentum, $M_V$ is the mass of the vector meson and $\epsilon^\mu$ the
polarization vector: $\epsilon^\mu(\pm1)=(0,~0,~\vec\epsilon_\perp),~
\epsilon^\mu(0)=\frac{1}{M_V}(\frac{-M_V^2}{P^+},~P^+,~0)$. Note that the longitudinal
polarization vector of the meson is not the same as that of the quark system due to $M_V\neq
M_0$.

It is straightforward to show that the decay constant of a pseudoscalar meson and a vector
meson can be represented by
 \be
 f_P&=&4\sqrt{\frac{3}{2}}\int\frac{dxd^2k_\perp}{2(2\pi)^3}
  \phi_P(x,k_\perp)\frac{{\cal A} }{\sqrt {{\cal A}^2+k_\perp^2}},\non\\
 f_V&=&4\sqrt {\frac{3}{2}}\int\frac{dxd^2k_\perp}{2(2\pi)^3}
  \frac{\phi_V(x,k_\perp)}{\sqrt {{\cal A}^2+k_\perp^2}}
   \frac{1}{M_{0}}\Big\{x(1-x)M_{0}^2+m_1m_2+k_\perp^2 \non\\
  &&+\frac{{\cal B}}{2W_V}\left[\frac{m_1^2+k_\perp^2}{1-x}
    -\frac{m_2^2+k_\perp^2}{x}-(1-2x)M_{0}^2\right]\Big\},
 \en
where
 \be
 &&{\cal A}=m_1x+m_2(1-x), \qquad {\cal B}=xm_1- (1-x)m_2,\non\\
 &&W_V=M_0+m_1+m_2.
 \en
In the above expression for the vector decay constant, we have used the plus component for the
longitudinal polarization vector. When the decay constants are known from the experimental
data, they can be used to constrain the parameters in the light cone wave functions.

\subsection{SCET light cone quark model}

Now, we discuss how to establish a light cone quark model utilizing soft collinear effective
theory. Since the $\bar B$ meson mass is dominated by the $b$ quark mass, the momentum
fraction for the spectator light antiquark $x$ is of order $\lqcd/m_b$. The variable $X\equiv
xm_b$ is of order of $\lqcd$ which is independent of $m_b$ in the limit $m_b\to\infty$. The B
meson wave function should have a scaling behavior in the heavy quark limit \cite{CZL}
 \be
  \phi_B(x,k_{\bot})\to \sqrt{m_b}\Phi(X,k_{\bot}),
 \en
where the factor $\sqrt{m_b}$ subtracts out the $m_b$ dependence of $\phi_B(x,k_{\bot})$ and
the function $\Phi(X,k_{\bot})$ is normalized as $\int dXd^2k_{\bot}|\Phi(X,k_{\bot})|^2=1$.
It is also found that $\Phi(X,k_{\bot})$ is a function of $v\cdot p_q$:
$\Phi(X,k_{\bot})\to\Phi(v\cdot p_q)$ with $p_q$ the momentum of spectator antiquark. This
observation is important in heavy-to-heavy transitions, however, because we work in the $\bar
B$ meson rest frame, it does not help to understand the \hl case. The light meson wave
function $\phi_M(x,k_{\bot})$ appeared in the \hl form factors is the wave function at
endpoint $x\sim \lqcd/E\to 0$ in the large energy limit. The form of light meson wave function
at endpoint is very important in determining the scaling behavior in $m_b$ of the \hl form
factors.

In the heavy quark limit, the heavy quark momentum is approximated as $p_b\cong m_b v$ with
other components  neglected. For the light energetic quark, $p^-\ll p_{\bot}\ll p^+$. Thus,
the light quark momentum $p$ is replaced by $p^{\mu}\cong (n_+\cdot p)\frac{n_-^{\mu}}{2}$. As
discussed before, in the soft collinear effective theory, the fields describing the heavy
quark is two component spinor $h_v$ and the energetic quark is spinor $\xi$. For our purpose,
we need the expression for the helicity sums for Dirac spinors in the heavy quark limit. For
heavy quark $h_v$, the leading order contribution is
 \be
  \sum\limits_{\lambda}h_v(\lambda)\bar h_v(\lambda)=(1+\vslash).
 \en
For light quark field $\xi$,  the helicity sum gives
 \be
  \sum\limits_{\lambda}\xi(p,\lambda)\bar \xi(p,\lambda)=
   \frac{\nslash_-}{2}.
 \en
The above two equations provide the spin symmetry relations for \hl form factors. While for
the spectator antiquark, it satisfies the relation given in Eq. (\ref{eq:spin}).

\begin{figure}[hbtp]
\vspace{0.4cm}
\begin{center}
\includegraphics[scale=0.8]{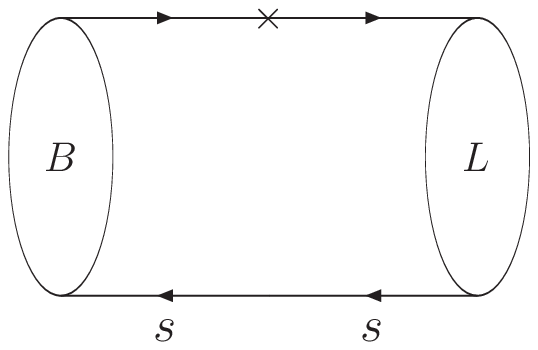}
\end{center}
\caption{The leading order contribution to \hl form factors with
``s" representing the soft momentum.} \label{Fig:hlsoft}
\end{figure}

The momenta for $\bar B$ and light meson are denoted by $P$ and $P'$, respectively. It is
convenient to work in the $\bar B$ meson rest frame and set $P'_{\bot}=0$. In this Lorentz
frame, the momentum transfer $q$ is purely longitudinal, i.e., $q_{\bot}=0$ and
$q^2=q^+q^-\geq 0$ covers the entire physical range.

The lowest order contribution to the  form factor comes from the soft Feynman diagram where
the spectator antiquark goes directly into the final light meson. The diagram is depicted in
Fig.~\ref{Fig:hlsoft}. The valence quark approximation guarantees that only the endpoint wave
function of the light meson overlaps with the $\bar B$ meson. We use $p_b$, $p_1$ and $p_q$ to
denote the momentum of the $b$ quark, the energetic quark and the spectator:
 \be
  && p_b^+=(1-x)P^+,    \qquad \qquad p_{b\perp}=-k_\perp, \non\\
  && p_1^+=(1-x')P'^+,  \qquad ~~~~   p_{1\perp}=-k_\perp, \non \\
  && p_q^+=xP^+=x'P'^+, \qquad ~~     p_{q\perp}=k_\perp,
 \en
where $P^+=M_B$ and $P'^+=2E$. The $x$, $x^\prime$ are the momentum fractions of the spectator
antiquark in $\bar B$ meson and in the final state meson, respectively. $x$ and $x^\prime$ are
connected by $x=x'r$. It is useful to define a variable $r\equiv P'^+/P^+=1-q^2/M_B^2$. Since
$x'$ varies from 0 to 1, thus $x$ varies from 0 to $r$.

Now, we are able to present the derivation of form factors in light cone approach with some
details. The $\bar B$ to pseudoscalar meson matrix element can be expressed as
 \be
  \la P|\bar \xi h_v|\bar B\ra = (-1)N_c\int^r_0 dx\int\frac{d^2k_\perp}
   {2(2\pi)^3}P^+\phi_P^*(x^\prime,k_\perp)\phi_B(x,k_\perp)
   \frac{P^+P'^+\sqrt{x(1-x)}\sqrt{x'(1-x')}}
   {\sqrt 2\tilde M_0\sqrt 2\tilde M_0'}
  \nonumber\\
  \times\mbox{Tr}\left[\frac{(\pslash_q-m_q)}{p_q^+}
   \frac{(i\gamma_5)}{\sqrt{N_c}}\frac{\nslash_-}{2}(1+\vslash)
   \frac{(-i\gamma_5)}{\sqrt{N_c}}\right],~~~~~~~~~~~~
 \en
where  $m_q$ is the mass of spectator antiquark. Since $x\sim x'\sim \lqcd/m_b$, we will
neglect $x,x'$ compared to $1$. The mass difference between $b$ quark mass and $\bar B$ meson
is neglected, i.e., $m_b\doteq M_B$. It is easy to obtain the relation $\sqrt{x(1-x)}\tilde
M_0=\sqrt{{\cal A}^2+k_{\bot}^2}$. Expanding the momentum and keeping the leading power
component, we get
 \be \label{eq:s1}
  \la P|\bar \xi h_v|\bar B\ra=\int^r_0 dx\int\frac{d^2k_\perp}{2(2\pi)^3}
  \frac{\phi_P^*(x^\prime,k_\perp)\phi_B(x,k_\perp)}
  {\sqrt{{\cal A}_B^2+k_\perp^2}\sqrt{{\cal A}_P^2+k_\perp^2}}
  ~ x m_b^2(p_q^- +m_q),
 \en
where $p_q^-=\frac{k_{\bot}^2+m_q^2}{xm_b}$. From Eqs. (\ref{eq:softFF}) and (\ref{eq:s1}),
one obtain
 \be \zeta_P=\frac{m_b}{2E}\int^r_0 dx \int\frac{d^2k_\perp}{2(2\pi)^3}
  \frac{\phi^*_P(x^\prime,k_\perp)\phi_B(x,k_\perp)}
  {\sqrt{{\cal A}_B^2+k_\perp^2}\sqrt{{\cal A}_P+k_\perp^2}}
  (xm_bm_q+m_q^2+k_\perp^2).
 \en
It shows that the leading order form factor $\zeta_P$ depends on the spectator quark mass
$m_q$, scaleless factor $m_b/E$ and non-perturbatively depends on $E$ through light meson wave
function $\phi_P(x',k_{\bot})$ at $x'\sim \lqcd/E$. The $m_b$ must be associated with $x$
means that the form factor depends on the non-perturbative scale $X=xm_b$ rather than hard
scale $m_b$ (except a normalization constant factor $\sqrt{m_b}$).

For $\bar B$ meson decays to longitudinal polarized vector, substituting the polarization
vector into the right hand side of eq.~(\ref{eq:softFF}), we get
 \be
  \langle V|\bar \xi\gamma_5 h_v|\bar B\rangle=-iM_V\zeta_{||}
   (\frac{P'^+}{M_V}-\frac{M_V}{P'^+})=-iP'^+\zeta_{||},
  \en
where we have dropped the sub-leading term. The expression in the
light cone approach gives
 \be
  \langle V|\bar \xi\gamma_5 h_v|\bar B\rangle&=&i
   \int_0^r dx\int \frac{d^2k_\perp}{2(2\pi)^3}
   \frac{\phi_V^*(x^\prime,k_\perp)\phi_B(x,k_\perp)}
   {2\sqrt{{\cal A}_B^2+k_\perp^2}
   \sqrt{{\cal A}_V^2+k_\perp^2}}x^\prime m_b^2\non\\
   \ \ \
   &&\times \mbox{Tr}\left[(\pslash_q-m_q)\left(-\hat\epsilon\!\!
   \slash+\frac{\hat\epsilon\cdot(p_1-p_q)}{W_V}\right)\frac{\nslash_-}{2}
   \gamma_5(1+\vslash)\gamma_5\right],\non\\
   \ \ \
  &=&\frac{-im_b^2P'^+}{2}\int^r_0  dx\int\frac{d^2k_\perp}{2(2\pi)^3}
  \frac{\phi_V^*(x^\prime,k_\perp)\phi_B(x,k_\perp)}
  {M_{0V}\sqrt{{\cal A}_B^2+k_\perp^2}\sqrt{{\cal A}_V^2+k_\perp^2}}\non\\
  &&\times x[2z^2(p_q^+
  +m_q)+\frac{p_q^-+z^2p_1^+}{W_V}(p_q^-+m_q)],
 \en
with $z\equiv M_{0V}/P'^+$. Although it seems that the first term is suppressed by
$\lambda=\sqrt{\Lambda_{QCD}/m_b}$, later we find that this term gives a relatively large
contribution in the numerical calculation. We obtain the expression for the longitudinal
leading order form factor as
 \be
  \zeta_{||}&=&\frac{m_b^2}{2}\int \frac {dx d^2k_\perp}{2(2\pi)^3}
  \frac{\phi_V^*(x^\prime, k_\perp)\phi_B(x,k_\perp)}
  {M_{0V}\sqrt{{\cal A}^2_B+k_\perp^2}\sqrt{{\cal A}_V^2+k_\perp^2}}\non\\
  &&\times x\left[2z^2(p_q^+ +m_q)+
   \frac{p_q^- +z^2 r m_b}{W_V}(p_q^- +m_q)\right].
 \en

Similarly, we can analyze the leading order transverse form factor. When performing the
calculation of $\zeta_\perp$, a formula for the transverse momentum integral is useful
 \be
 \int d^2k_\perp (\epsilon \cdot p_{1})p_q^\alpha=
  \frac{1}{2}\int d^2k_\perp k_\perp^2\epsilon^\alpha.
 \en
The expression for $\bar B$ to transversely polarized vector meson is
 \be
  \la V|\bar \xi\gamma_{\bot}^{\mu} h_v|\bar B\ra&=&i
   \int_0^r dx\int \frac{d^2k_\perp}{2(2\pi)^3}
   \frac{\phi_V^*(x^\prime,k_\perp)\phi_B(x,k_\perp)}
   {2\sqrt{{\cal A}_B^2+k_\perp^2}
   \sqrt{{\cal A}_V^2+k_\perp^2}}x m_b^2\non\\
   &&\times \mbox{Tr}\left[(\pslash_q-m_q)\left(-\hat\epsilon\!\!
   \slash+\frac{\hat\epsilon\cdot(p_1-p_q)}{W_V}\right)\frac{\nslash_-}{2}
   \gamma_{\bot}^{\mu}(1+\vslash)\gamma_5\right],\non\\
  &=&-m_b^2\int\frac{dx d^2k_\perp}{2(2\pi)^3}
   \frac{\phi_V^*(x^\prime,k_\perp)\phi_B(x,k_\perp)}
   {\sqrt{{\cal A}_B^2+k_\perp^2}\sqrt{{\cal A}_V^2+k_\perp^2}}
   \epsilon^{\mu\nu\rho\sigma}\epsilon_\nu^* v_\rho n_{-\sigma}
   ~x(p_q^-+m_q+\frac{k_\perp^2}{W_V}).\non\\
 \en

It is straightforward to get:
 \be
 \zeta_\perp&=&\frac{m_b^2}{2E}\int\frac{dx d^2k_\perp}{2(2\pi)^3}
  \frac{\phi_V^*(x^\prime,k_\perp)\phi_B(x,k_\perp)}
  {\sqrt{{\cal A}_B^2+k_\perp^2}\sqrt{{\cal A}_V^2+k_\perp^2}}
  ~ x(p_q^-+m_q+\frac{k_\perp^2}{W_V}).
 \en


\subsection{Higher order corrections to the heavy-to-light form
factors in the light cone perturbation theory}

\begin{figure}[hbtp]
\vspace{0.4cm}
\begin{center}
\includegraphics[scale=0.8]{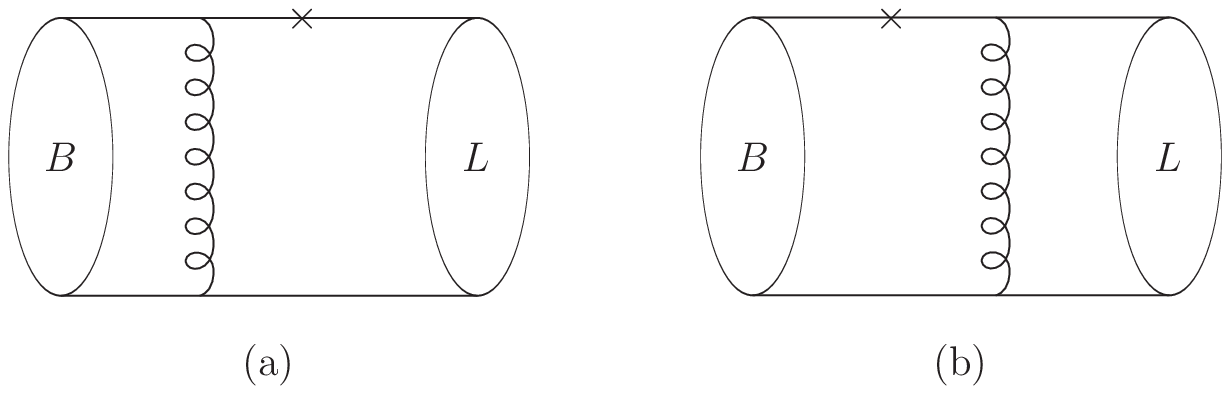}
\end{center}
\caption{The one gluon exchange contributions to \hl form factors
with the signs ``$\times$'' representing electro-weak vertex. }
\label{Fig:hloneg}
\end{figure}

In this subsection, we will derive the higher order corrections for the heavy to light form
factors in the light cone perturbation theory of QCD. Besides the leading order soft
contributions to the universal form factors, the next-to-leading order contribution is the
kind of diagrams shown in Fig. \ref{Fig:hloneg} with one hard gluon exchange (about the vertex
corrections, see \cite{BF1, SCETff1}).

A four-component Dirac field $\psi$ can be decomposed into two-component spinors $\xi$ and
$\eta$ by
 \be
  \psi=\xi+\eta, \qquad
  \xi \equiv P_-\psi=\frac{\nslash_-\nslash_+}{4}\psi, \qquad
  \eta\equiv P_+\psi=\frac{\nslash_+\nslash_-}{4}\psi,
 \en
with equations of motion for spinors $\xi$ and $\eta$ are
 \be
 in_-\cdot D\frac{\nslash_+}{2}~\xi+(i\dslash_{\bot}-m)\eta=0;
  \label{eq:xi}\\
 in_+\cdot D\frac{\nslash_-}{2}~\eta+(i\dslash_{\bot}-m)\xi=0.
  \label{eq:eta}
 \en
In \lc quantization, the time variable is chosen to be different from the conventional one
$t=x^3$. We adopt the light cone time as $\tau=n_+\cdot x$ and then the time-like derivative
is $n_-\cdot \partial$. In Eq. (\ref{eq:eta}), there is no time derivative. Thus $\eta$ is a
constrained field\footnote{In some references, $\xi$ is called ``good" component and $\eta$ is
called ``bad" component.}, since it is determined by $\xi$ at any time of $n_+\cdot x$. From
Eq. (\ref{eq:eta}), $\eta$ field is obtained as
 \be
 \eta=\frac{1}{in_+\cdot
 D}(i\dslash_{\bot}+m)\frac{\nslash_+}{2}\xi.
 \en

For the gluon field, it satisfies the color Maxwell equation
$\partial_{\mu}F^{a\mu\nu}=gJ^{a\nu}$ where $J^{a\nu}$ is the quark current. By using the
constraint $n_+\cdot A=0$, we obtain one relation $(n_+\cdot
\partial)\partial_{\mu}A^{a\mu}=-g(n_+\cdot J^a)$. Thus, the field component
$n_-\cdot A$ is not a dynamical variable but determined by $A_{\bot}$ through
 \be
  n_-\cdot A=\frac{2}{n_+\cdot \partial}\partial_{\bot}
   \cdot A_{\bot}-\frac{2}{(n_+\cdot \partial)^2}g(n_+\cdot J^a).
 \en

The Feynman rules for $\xi$ and $A_{\bot}$ have been derived, such as in \cite{SB} which are
not useful for our purpose. We prefer to use another formulation given in \cite{LB}. In \lc
perturbation theory, the diagrams are $n_+\cdot x$-ordered and all particles are on
mass-shell. For the propagator of quark, it contains an instantaneous part, in particular
 \be \label{eq:lc1}
 \frac{i(\pslash+m)}{p^2-m^2+i\epsilon}=
  \frac{i(\pslash_{on}+m)}{p^2-m^2+i\epsilon}
  +\frac{i~\nslash_+}{2n_+\cdot p},
 \en
where $p_{on}$ is the on-shell momentum $p_{on}=(n_-\cdot p, \frac{p_{\bot}^2+m^2}{n_-\cdot
p},p_{\bot})$ and $p_{on}^2=m^2$. The second term in the quark propagator
$\frac{\nslash_+}{2n_+\cdot p}$ is the instantaneous part induced by integrating out the field
$\eta$. For the gluon field, the polarization sum is written as
 \be \label{eq:lc2}
 d_{\mu\nu}(k)\equiv \sum\limits_{\lambda_g}\epsilon_{\mu}(k,\lambda_g)
  \epsilon_{\nu}^*(k,\lambda_g)=\sum\limits_{i=1,2}
  \left [-n_{+\mu}\frac{\epsilon^{(i)}\cdot k}{n_+\cdot k}
   +\epsilon^{(i)}_{\mu}\right ]
  \left [-n_{+\nu}\frac{\epsilon^{(i)}\cdot k}{n_+\cdot k}
   +\epsilon^{(i)}_{\nu} \right ],
 \en
where $\epsilon^{(i)}$ are purely transverse vectors: $\epsilon^{(i)+}=\epsilon^{(i)-}=0$ and
$\epsilon^{(i)*}_\bot \cdot \epsilon^{(j)}_\bot=\delta^{ij}$. There are two terms in the
bracket of Eq.~(\ref{eq:lc2}): the first term $n_{+\mu}\frac{\epsilon^{(i)}\cdot k}{n_+\cdot
k}$ comes from the longitudinal component $n_+\cdot A$ and the second $\epsilon^{(i)}_{\mu}$
from the transverse component $A_{\bot}$. If the gluon momentum is chosen to be in the
longitudinal direction, then $\epsilon^{(i)}\cdot k=0$ and only the transverse components
$\epsilon^{(i)}$ are remained. It reflects the fact that the physical gluon is transverse
polarized. In the above rules, the choice of $n_+$ and $n_-$ is arbitrary and there is a
symmetry by exchanging them. In this way, we obtain the light cone quantization rules for the
light cone time $n_-\cdot x$.

For the one gluon exchange diagram given in Fig.~\ref{Fig:hloneg}, the amplitude at the quark
level is given in the conventional covariant form as
 \be \label{eq:A}
 A=\frac{g^2}{k^2}d_{\mu\nu}\left\{\bar u(p_1)\Gamma
   \frac{(\pslash_{q1}+m_b)}{p_{q1}^2-m_b^2} T^A\gamma^{\mu}
   b(p_b)~\bar v_s(p_q)T^A\gamma^{\nu}v(p_2)~~~ \right. \non\\
   \left. +\bar u(p_1)T^A\gamma^{\mu}\frac{\pslash_{q2}}
   {p_{q2}^2}\Gamma b(p_b)~\bar v_s(p_q)
   T^A\gamma^{\nu}v(p_2)\right\},
 \en
where $u(v)$ are light quark (antiquark) spinor, $b(v_s)$ are $b$ quark (spectator antiquark)
spinor; $p_{q1, q2}$ are the internal quark momenta, $k$ the exchanged gluon momentum, and
$k=p_2-p_q$, $p_{q1}=p_b-p_2$, $p_{q2}=p_1+p_2-p_q$. The first term of the amplitude comes
from contribution of Fig.~\ref{Fig:hloneg}(a) and the second term from the diagram
Fig.~\ref{Fig:hloneg}(b). We have neglected the light quark masses. For the second term in
Eq.~(\ref{eq:A}), we use the light cone quantization rules of
Eqs.~(\ref{eq:lc1},\ref{eq:lc2}). While for the first term in Eq.~(\ref{eq:A}), the exchanged
rules of Eqs.~(\ref{eq:lc1},\ref{eq:lc2}) by $n_-\leftrightarrow n_+$ is applied. Thus, the
amplitude is rewritten in the light cone form by
 \be \label{eq:lc3}
 A=\frac{g^2}{k^2}\left\{\bar u\Gamma\left[\frac{\nslash_-}
   {2n_-\cdot p_{q1}}+\frac{(\pslash_{q1})_{on}+m_b}{p_{q1}^2-m_b^2}
   \right]T^A \left[\gamma_{\bot}^{\mu}-\frac{\nslash_-}
   {n_-\cdot k}k_{\bot}^{\mu}\right]
   b~\bar v_s T^A\left[\gamma_{\bot\mu}-\frac{\nslash_-}{n_-\cdot k}
   k_{\bot\mu}\right]v \right.  \non \\
  +\left.\bar u T^A\left[\gamma_{\bot}^{\mu}-
   \frac{\nslash_+}{n_+\cdot k}k_{\bot}^{\mu}\right]\left[
   \frac{\nslash_+}{2n_+\cdot p_{q2}}+\frac{(\pslash_{q2})_{on}}
   {p_{q2}^2}\right]\Gamma b~\bar v_s T^A\left[\gamma_{\bot\mu}-
   \frac{\nslash_+}{n_+\cdot k}k_{\bot\mu}\right]v \right\}.
  \en

Neglecting the contributions suppressed by $\lqcd/m_b$, we find the contribution from the
instantaneous interaction part is
 \be \label{eq:Ah}
 A^h=\frac{-g^2}{(n_+\cdot p_2)(n_-\cdot p_q)}\left\{\bar \xi_{n_-}
   \Gamma\frac{\nslash_-}{2m_b}T^A \gamma_{\bot}^{\mu} h_v~\bar v_s
   T^A \gamma_{\bot\mu}\xi_{n_-}
   +\bar \xi_{n_-}T^A\gamma_{\bot}^{\mu}\frac{\nslash_+}
   {2n_+\cdot P'}\Gamma h_v~\bar v_s
   T^A\gamma_{\bot\mu}\xi_{n_-}\right\}.
 \en
This contribution is not singular for the leading twist distribution amplitudes of $B$  and
light mesons. It is usually called ``hard" contribution which breaks the spin symmetry due to
$\nslash_-$ and $\nslash_+$ matrices. In the light cone language, the hard gluon exchange
contributions come from the instantaneous quark interactions and the transversely polarized
gluons.
The hard one gluon exchange contributions can not be absorbed into
the three universal form factors because this type higher order
contributions break the spin symmetry in the leading order.

\section{Numerical Results and Discussions}\label{section3}

The physical \hl form factors contain both hard and soft contributions.  In this study, we
concentrate on the leading order soft form factors. The next-to-leading order $\alpha_s$
corrections, which breaks the spin symmetry, will be calculated in a future work. In order to
obtain the numerical results, we have to determine the wave functions of the hadrons which
contain all information of the hadron state. The full solution needs great efforts, so we  use
the phenomenological Gaussian-type wave function:
 \be
  \phi(x,k_\perp)=N\sqrt{\frac{dk_z}{dx}}\mbox{exp}
  (-\frac{\vec k^2}{2\omega^2}),
 \en
where $N=4(\pi/\omega^2)^{3/4}$ and $k_z$ of the internal momentum $\vec k=(\vec k_\perp,k_z)$
is defined through
 \be
  1-x=\frac{e_1-k_z}{e_1+e_2}, \qquad
  x=\frac{e_2+k_z}{e_1+e_2},
  \en
with $e_i=\sqrt {m_i^2+\vec k_i^2}=\frac{x_i M_0}{2}+
\frac{m_i^2+k_{\perp i}^2}{2x_i M_0}$. We
then have
 \be
  k_z=\frac{xM_0}{2}-\frac{m_2^2+k_\perp^2}{2xM_0},\qquad
  \frac{dk_z}{dx}=\frac{e_1e_2}{x(1-x)M_0}.
 \en

In this wave function, the distribution of the momentum is determined by the quark mass and
the parameter $\omega$. The quarks are constituent quarks and the quark masses are usually
chosen as:
 \be
  m_{u,d}=0.25 {\mbox { GeV}}, \qquad
  m_s=0.40{\mbox { GeV}},      \qquad
  m_b=4.8 \mbox{ GeV}.
 \en
The parameter $\omega$ can be determined by the hadronic results, for example, the decay
constants \cite{flavor}.

As for the decay constants of $\eta$ and $\eta^\prime$, we should pay much more attention on
the mixing of these two particles. Although the quark model has achieved great successes, we
still don't have the definite answer on the exact components of these two mesons. The study of
$B$ to $\eta^{(\prime)}$ decays, especially the study on  $B \to \eta^{(\prime)}$ form factor,
can help us to understand their intrinsic characters (For a recent study, please see
\cite{CKL}). Here we view these two particles as the conventional two quark states. As for the
mixing, we use the quark flavor basis proposed by Feldmann and Kroll \cite{FKmixing}, i.e.
these two mesons are made of $\eta_n=\bar nn=(\bar uu+\bar dd)/\sqrt 2$ and $\eta_s=\bar ss$:
\begin{equation}
   \left( \begin{array}{c}
    |\eta\rangle \\ |\eta'\rangle
   \end{array} \right)
   = U(\phi)
   \left( \begin{array}{c}
    |\eta_n\rangle \\ |\eta_s\rangle
   \end{array} \right) \;,
\end{equation}
with the mixing matrix,
\begin{equation}
U(\phi)=\left( \begin{array}{cc}
    \cos\phi & ~-\sin\phi \\
    \sin\phi & \phantom{~-}\cos\phi
   \end{array} \right)\;,
\end{equation} where $\phi$ is the mixing angle.
In this mixing scheme, only two decay constants $f_n(n=u,d)$ and
$f_s$ are needed:
\begin{eqnarray}
   \langle 0|\bar n\gamma^\mu\gamma_5 n|\eta_n(P)\rangle
   &=& \frac{i}{\sqrt2}\,f_n\,P^\mu \;,\nonumber \\
   \langle 0|\bar s\gamma^\mu\gamma_5 s|\eta_s(P)\rangle
   &=& i f_s\,P^\mu \;.\label{deffq}
\end{eqnarray}
This   is based on the assumption that the intrinsic $\bar nn(\bar ss)$ component is absent in
the $\eta_s(\eta_n)$ meson, i.e., based on the $OZI$ suppression rule. These decay constants
have been determined from the related exclusive processes as \cite{FKmixing}:
 \be
  f_n=(1.07\pm0.02)f_\pi, \qquad
  f_s=(1.34\pm0.06)f_\pi.
 \en
In the following we will calculate the form factors of $B\to \eta_n$ and $B_s\to \eta_s$. The
gluonic contribution to $B\to \eta^{(\prime)}$ has also been studied in Ref.~\cite{CKL}. We
will neglect it as it is very small.

We use the following results for the decay constants as input in the light front wave
functions:
 \be
   f_B=0.190 \mbox{ GeV},~   &
     f_{B_s}=0.236 \mbox{GeV},~~
     f_\pi=0.132 \mbox{ GeV},~ &
     f_K=0.160 \mbox{ GeV}, ~\non \\
 f_\rho=0.205{\mbox{ GeV}}, ~  &
     f_\omega=0.195\mbox{GeV}, ~~~~
     f_{K^\star}=0.217\mbox{ GeV},~&
     f_\phi=0.231\mbox{ GeV}.
 \en
Then the parameters $\omega$ in the light-front wave functions are
determined from these decay constants as:
 \be
  && \omega_B=0.55^{+0.05}_{-0.04}\mbox{GeV},\qquad \qquad
     \omega_{B_s}=0.64^{+0.05}_{-0.06}\mbox{GeV}, \non\\
  && \omega_\pi=0.33\mbox{GeV},\qquad \qquad \qquad
     \omega_K=0.38\mbox{GeV}, \non\\
  && \omega_n=0.38^{+0.09}_{-0.08}\mbox{GeV}, \qquad \qquad
     \omega_s=0.39^{+0.06}_{-0.06}\mbox{GeV}, \non \\
  && \omega_\rho=0.31^{+0.03}_{-0.03}\mbox{GeV},\qquad \qquad
     \omega_\omega=0.29^{+0.03}_{-0.03}\mbox{GeV}, \non\\
  && \omega_{K^\star}=0.33^{+0.03}_{-0.03}\mbox{GeV},~~~ \qquad \
     \omega_\phi=0.35^{+0.03}_{-0.03}\mbox{GeV},
 \en
where  the uncertainties come from  varying the decay constants of the heavy and the light
mesons by $10\%$. Some light meson decay constants have been determined to a high accuracy,
for example, $f_{\pi}$, $f_K$. We neglect the uncertainties for them.

\subsection{Results for $B\to P$ form factor $\zeta_P$}

Now we  are ready to give the numerical results of the $B$ to pseudoscalar soft form factors
at $q^2=0$, i.e. $E=m_B/2$. Using the above parameters, we obtain the results as follows:
 \begin{eqnarray}
 && \zeta_P^{B\to \pi}(\frac{m_B}{2})=
      0.247,~ \qquad\qquad
    \zeta_P^{B\to K}(\frac{m_B}{2})=
      0.297,\non\\
 && \zeta_P^{B\to \eta_n}(\frac{m_B}{2})=
      0.287^{+0.059}_{-0.065},\non\\
 && \zeta_P^{B_s\to K}(\frac{m_B}{2})=
      0.290, \qquad \qquad
    \zeta_P^{B_s\to \eta_s}(\frac{m_B}{2})=
      0.288^{+0.047}_{-0.052},\label{zetapB}
 \end{eqnarray}
where the uncertainties are from the decay constant of the light mesons. We also find  the
uncertainties caused by $B$ meson decay constants are rather small and thus we neglect these
uncertainties. In Ref.~\cite{SCETLCSR1}, the SCET sum rule result is calculated as
$\zeta_P^{B\to \pi}=0.27$ which is consistent with our result within theoretical errors. The
physical form factors can be obtained directly using the relation in Eq.~(\ref{reduction}). At
maximally recoil $r=1$, $f_+$ and $f_0$ are equal to each other, which are exactly the soft
form factor $\zeta_P$; $f_T$ is slightly larger. Table \ref{physicalformPP} lists the $B\to P$
form factors at $q^2=0$.

\begin{figure}[[hbtp]
\vspace{-0.2cm}
\begin{center}
\includegraphics[scale=0.4]{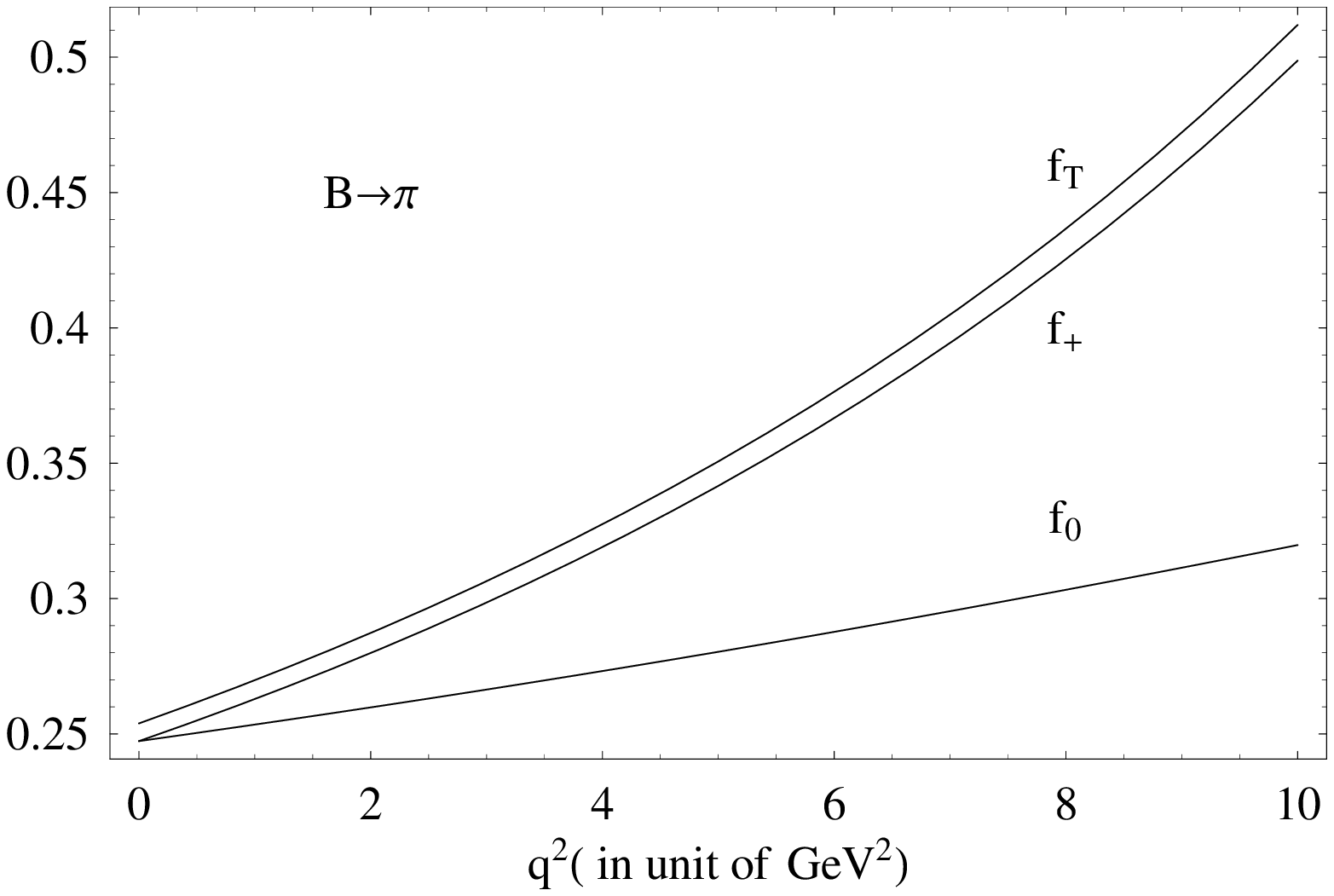}
\includegraphics[scale=0.4]{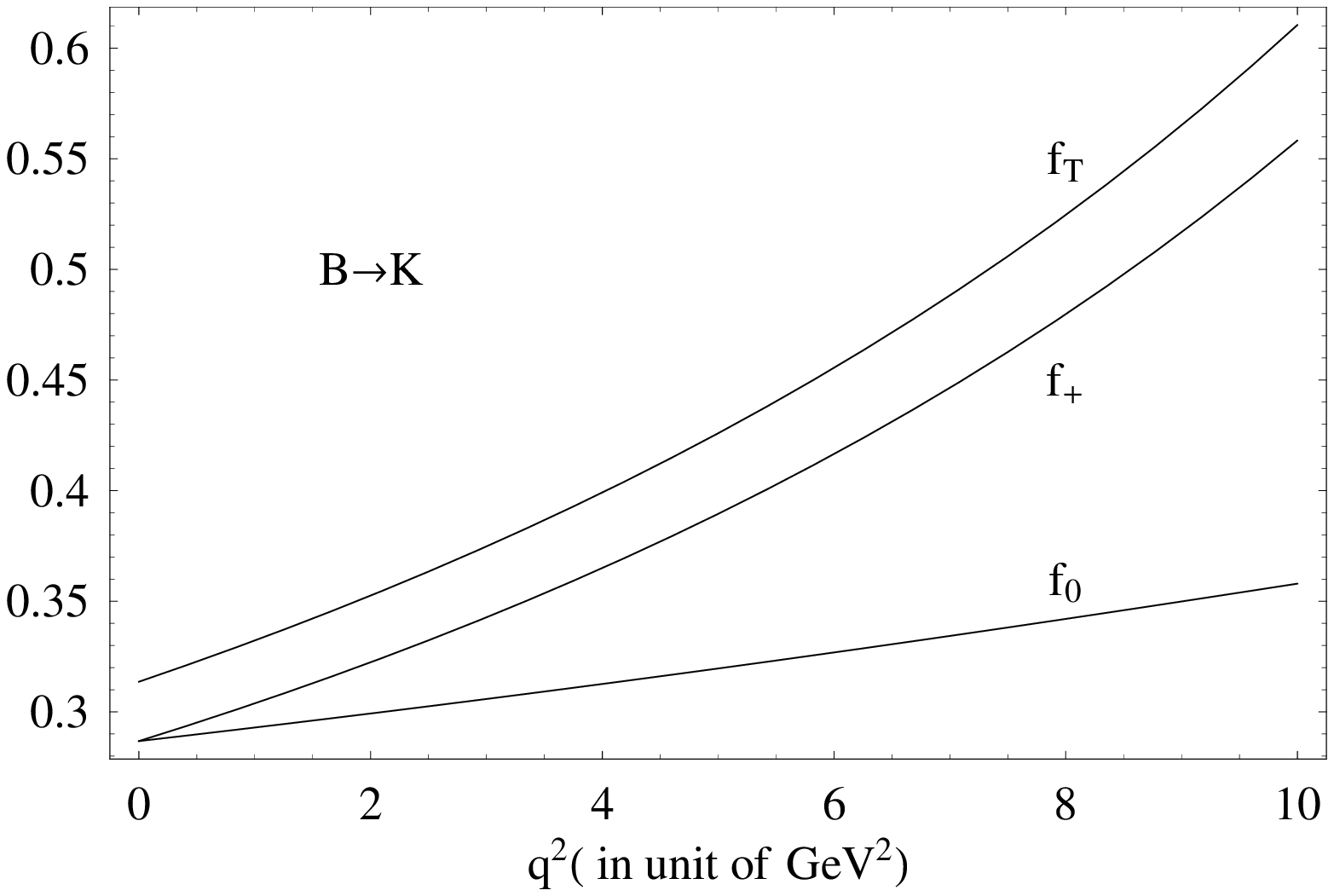}
\includegraphics[scale=0.4]{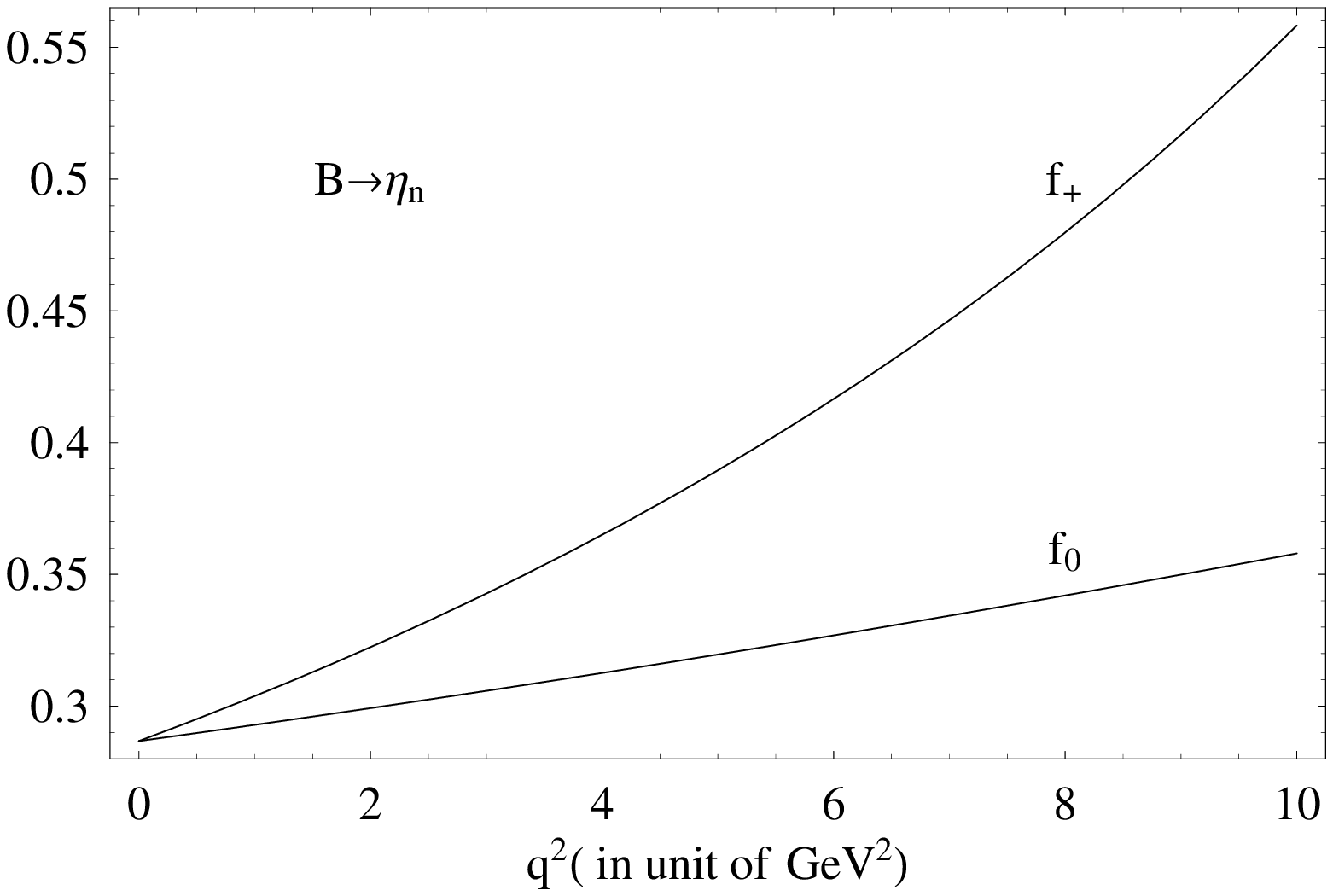}
\includegraphics[scale=0.4]{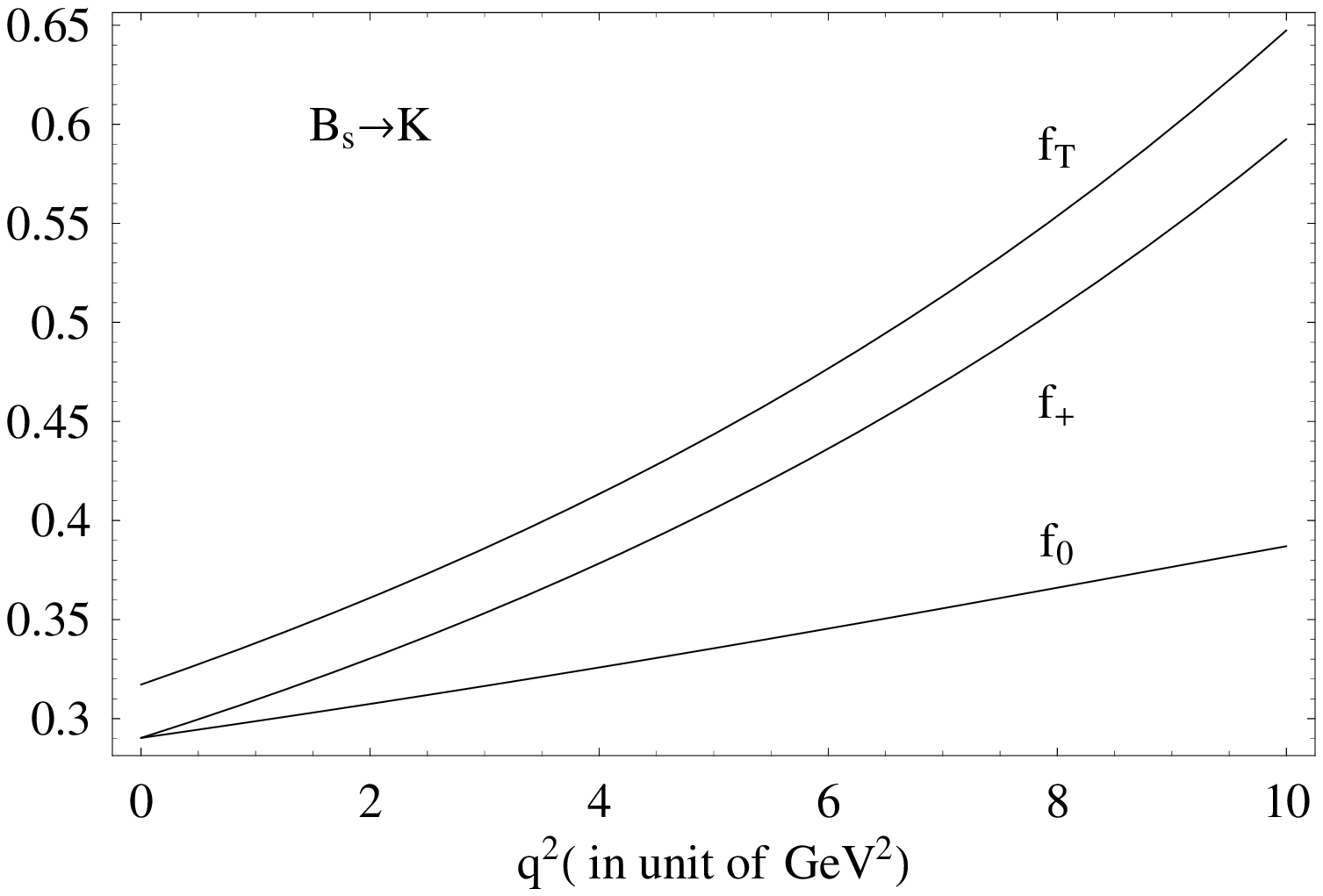}
\includegraphics[scale=0.4]{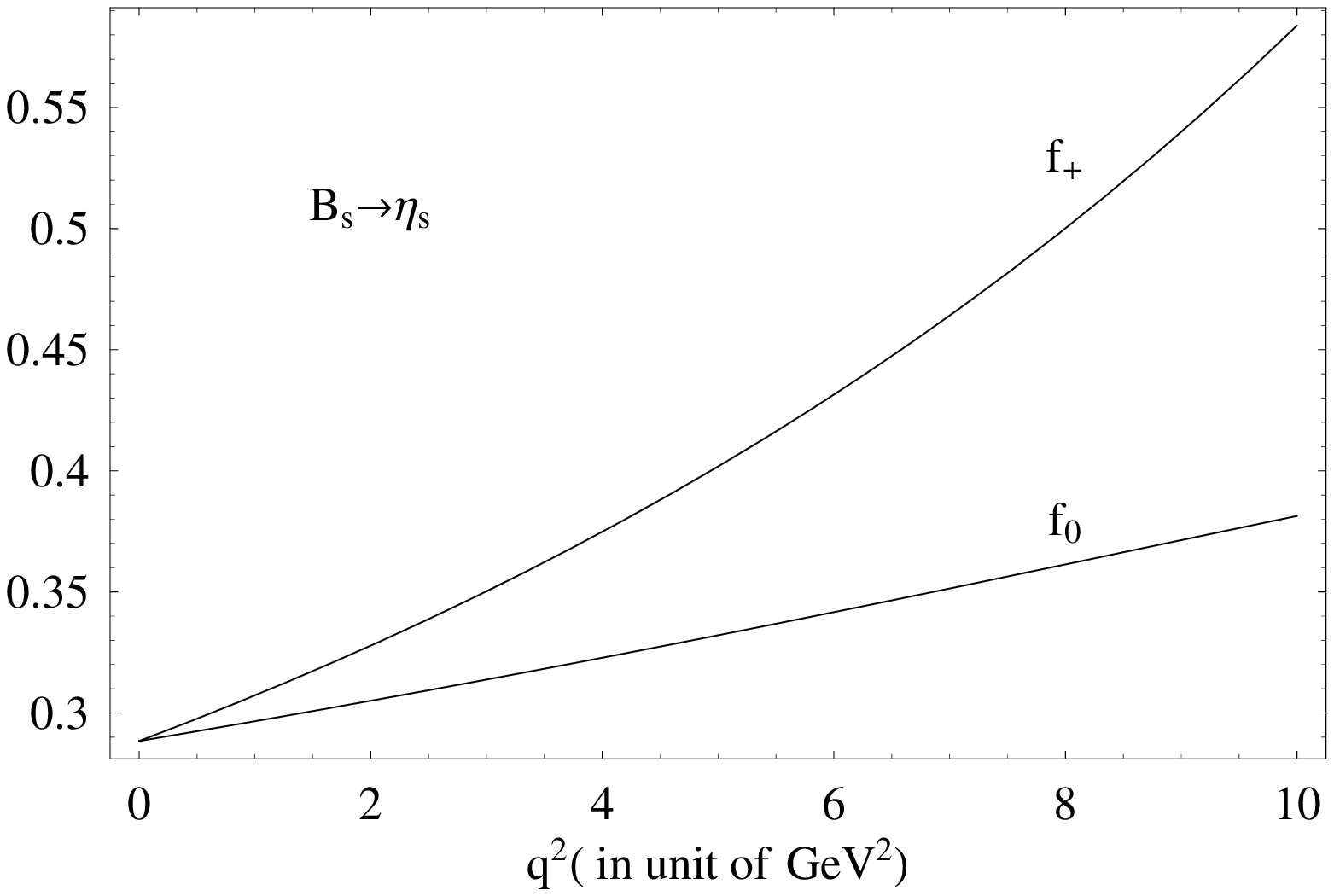}
\end{center}
\vspace{-0.2cm} \caption{{The $q^2$ dependence of $B_{(s)}\to P$
form factors. In this figure, we plot $f_+$, $f_0$ and $f_T$ for
$B\to\pi$, $B\to K$ and $B_s\to K$ transitions. But for $B\to\eta_n$
and $B_s\to \eta_s$, only the first two form factors are shown for
the ambiguity of the mass for $\eta_n$ and $\eta_s$ in
$f_T$.}}\label{BPqdepend}
\end{figure}

\begin{table}\caption{The physical $B_{(u,d,s)}\to P$ form factors at maximally recoil
using the usual LCQM \cite{CCH2}, LCSR \cite{BZpseudo}, LQCD
\cite{LQCD1,LQCD2,LQCD3} and PQCD \cite{PQCD} approaches.  The
different values for $f_+(B\to\pi)$ in Ref.~\cite{LQCD2}
correspond to different extrapolations. }
\begin{center}
\begin{tabular}{cc|c|c|c|c|c|c|c}
\hline \hline && LCQM\cite{CCH2} & LCSR\cite{BZpseudo}&PQCD
\cite{PQCD}& LQCD\cite{LQCD1} & LQCD
\cite{LQCD2}&LQCD\cite{LQCD3}& This work
\\ \hline
 $B\to\pi$& $f_+$  &$0.25$   &$0.258$  &$0.292$   &$0.27$  &$0.27(0.26)$ &$0.23$ &$0.247$ \\
\ \ \     & $f_T$  &         &$0.253$  &$0.278$   &        &             &       &$0.253$ \\
          & $f_0$  &$0.25$   &         &$0.292$   &$0.27$  &             &       &$0.247$\\ \hline
$B\to K$  & $f_+$  &$0.35$   &$0.331$  &$0.321$   &        &             &       &$0.297$  \\
\ \ \     & $f_T$  &         &$0.358$  &$0.311$   &        &             &       &$0.325$  \\
\ \ \     & $f_0$  &$0.35$   &         &$0.321$   &        &             &       &$0.297$ \\
\hline $B\to\eta_n$\footnote{The form factors of $B\to\eta$ is
calculated in LCSR rather than that of $B\to\eta_n$}
          & $f_+$  &         &$0.275$   &         &        &             &       &$0.287$  \\
\ \ \     & $f_T$  &         &$0.285$   &         &        &             &       &  \\
\ \ \     & $f_0$  &         &          &         &        &             &       &$0.287$ \\
\hline
$B_s\to K$& $f_+$  &         &          &         &        &             &       &$0.290$  \\
\ \ \     & $f_T$  &         &          &         &        &             &       &$0.317$  \\
\ \ \     & $f_0$  &         &          &         &        &             &       &$0.290$ \\
\hline
$B_s\to\eta_s$ & $f_+$&      &          &         &        &             &       &$0.288$    \\
            & $f_0$ &        &          &         &        &             &       &$0.288$ \\
 \hline \hline
\end{tabular}\label{physicalformPP}
\end{center}
\end{table}

These $B\to P$ form factors have also been studied systemically in the usual light-cone quark
model \cite{CCH1,CJK,CCH2}, the light cone sum rules \cite{BZpseudo} and PQCD approach
\cite{PQCD}. Although Lattice QCD cannot give direct predictions on the $B$ to light form
factors at large recoiling, there are some studies using the extrapolations from the results
at large $q^2$: in quenched LQCD \cite{LQCD1} and in unquenched LQCD \cite{LQCD2,LQCD3}. We
cite these results in Table~\ref{physicalformPP}.

Comparing the results in Table~\ref{physicalformPP}, we can find that our leading-order
results agree with the results calculated using other approaches. The numerical results of
higher order corrections which should be small in our approaches will be taken into account in
future work.

We compare our approach with the previous light cone quark models. As in the conventional form
of \cite{CCH1} where the quarks are on-shell, the calculation of form factors are in the
physical momentum regions $q^2\geq 0$. The difference between the approach in \cite{CCH1} and
ours is that we make approximations in the heavy quark mass and large energy limit. The
consistency of the numerical predictions in the two methods means that our result is the
leading dominant contribution. In the covariant form in \cite{Jaus2}, the quarks are
off-shell. The evaluations are performed in the momentum regions $q^2<0$ and the analytic
continuation is required to obtain the physical form factors. The advantage of this approach
is that the zero-mode $(k^+=0)$ contribution does not occur. In our method, the zero-mode
contribution vanish in the heavy quark mass and large energy limit.

Since our analysis is within the SCET framework, we should make sure that the final state
meson is energetic. The energy of the light meson should be larger than $\sqrt
{m_B\Lambda_{QCD}}\sim 1.5 \mbox{ GeV}$ in order to ensure it as a collinear meson. From this
constraint, we can get $q^2=m_B^2-2m_BE<10\mbox{ GeV}^2 $. Thus we can directly calculate the
form factors in the range of $0<q^2<10\mbox{ GeV}^2$ and the results should be reliable. We
plot the $q^2$ dependence of the $B_{(s)} \to P$ form factors in Fig.~\ref{BPqdepend}. In this
figure, the form factors $f_+(q^2)=\zeta_P(E)$, $f_T(q^2)=\frac{M_B+M_P}{M_B}\zeta_P(E)$ and
$f_0(q^2)=\frac{2E}{m_B}\zeta_P(E)$ are plotted. The $q^2$ dependence of $f_+$ and $f_T$ are
essentially the same except the only difference of the form factor at $q^2=0$.  The curve of
$f_0(q^2)$ is more flat than the other two because of the compensation of the factor
$r=\frac{2E}{m_B}=1-\frac{q^2}{m_B^2}$.

In order to study the analytic $q^2$ dependence of the results for the form factors, we fit
the data by adopting the simple parametrization: \be
f(q^2)=\frac{f(0)}{1-a(q^2/m_B^2)+b(q^2/m_B^2)^2},\en where $f(0)$ are the results at $q^2=0$
which have been discussed as above, while $a$ and $b$ are the parameters. The fitted results
for these two parameters are summarized in Table~\ref{zetap}. From Fig.~\ref{BPqdepend}, we
can see that all of the curves are close to be a straight line and the parameters $b$ should
be rather small. The results from the parametrization also verify this expectation. Our
results for parameters $a$ for different processes are also close to each other: around
$a=1.5$ for $f_+$ and $f_T$ or $a=0.5$ for $f_0$.

\begin{table}\caption{The parameters in the parametrization of
$B_{(u,d,s)}\to P$ form factors. The fitted values of $a$ and $b$
for $f_T$ are the same with the ones in $f_+$.}
\begin{center}
\begin{tabular}{c c|c|c|c|c|c}
\hline \hline
      &   & $f_+^{B\to\pi}$&$f_+^{B\to K}$&$f_+^{B\to \eta_n}$
      &$f_+^{B_s\to K}$&$f_+^{B_s\to \eta_s}$\\ \hline
 $a$  &   & $1.43$         & $1.28$       &  $1.31$      & $1.51$     & $1.49$   \\ \hline
 $b$  &   & $0.08$         & $0.00$       &  $-0.00$     & $0.23$     & $0.22$   \\ \hline \hline
      &   & $f_0^{B\to\pi}$&$f_0^{B\to K}$&$f_0^{B\to \eta_n}$
      & $f_0^{B_s\to K}$& $f_0^{B_s\to \eta_s}$\\ \hline
 $a$  &   & $0.56$         &  $ 0.46$     &  $0.48$      & $0.66$    & $0.64$  \\ \hline
 $b$  &   & $-0.14$        &  $-0.08$     &  $-0.14$     & $-0.00$   & 0.00    \\ \hline \hline
\end{tabular}\label{zetap}
\end{center}
\end{table}

\subsection{Results for $B\to V$ form factors }

Similar analysis can also be applied to $B\to V$ form factors. At
$q^2=0$, the results for the $B\to V$ soft form factors are
 \begin{eqnarray}
  && \zeta_{||}^{B\to\rho}(\frac{m_B}{2})=0.260^{+0.028}_{-0.030},~~\qquad
     \zeta_\perp^{B\to\rho}(\frac{m_B}{2})=0.260^{+0.030}_{-0.031},\non\\
  && \zeta_{||}^{B\to\omega}(\frac{m_B}{2})=0.240^{+0.029}_{-0.031},~~\qquad
     \zeta_\perp^{B\to\omega}(\frac{m_B}{2})=0.239^{+0.031}_{-0.031},\non\\
  && \zeta_{||}^{B\to K^*}(\frac{m_B}{2})=0.284^{+0.025}_{-0.027},\qquad
     \zeta_\perp^{B\to K^*}(\frac{m_B}{2})=0.290^{+0.027}_{-0.029},\non\\
  && \zeta_{||}^{B_s\to K^*}(\frac{m_B}{2})=0.279^{+0.030}_{-0.030},\qquad
     \zeta_\perp^{B_s\to K^*}(\frac{m_B}{2})=0.271^{+0.030}_{-0.030},\non\\
  && \zeta_{||}^{B_s\to\phi}(\frac{m_B}{2})=0.279^{+0.029}_{-0.030},~~\qquad
     \zeta_{\perp}^{B_s\to\phi}(\frac{m_B}{2})=0.276^{+0.030}_{-0.030},
 \end{eqnarray}
where the uncertainties are from the decay constants of the light mesons. In order to make a
comparison, we collect the results for the physical form factors in LCQM \cite{CCH1,CCH2},
LCSR \cite{BZvector}, PQCD \cite{PQCD} approach, LQCD \cite{LQCD1,LQCD4} and our leading-order
results in Table~\ref{physicalformPV}. Our results are consistent with other approaches except
for the smaller $T_{2,3}$ and larger $T_1$ in PQCD approaches.

\begin{table}\caption{The physical $B\to V$ form factors at maximally recoil,
  i.e. $q^2=0$. }
\begin{center}
\begin{tabular}{cc|c|c|c|c|c}
\hline \hline
         &       & $B\to\rho$ &$B\to K^*$     &$B\to\omega$ &$B_s\to K^*$  &$B_s\to\phi$ \\
 \hline
LCQM\cite{CCH2}&$V$ &$0.27$      &$ 0.31$        &             &              &        \\
\ \ \      & $A_0$  &$0.28$      &$0.31$         &             &              &        \\
\ \ \      & $A_1$  &$0.22$      &$0.26$         &             &              &        \\
\ \ \      & $A_2$  &$0.20$      &$ 0.24$        &             &              &        \\
\hline
LCSR\cite{BZvector}&$V$&$0.323$  &$0.411$        &$0.293$      &$0.311$       &$0.434$ \\
\ \ \      & $A_0$  &$0.303$     &$0.374$        &$0.281$      &$0.360$       &$0.474$ \\
\ \ \      & $A_1$  &$0.242$     &$0.292$        & $0.219$     &$0.233$       &$0.311$ \\
\ \ \      & $A_2$  &$0.221$     &$0.259$        &$0.198$      &$0.181$       &$0.234$ \\
\ \ \      & $T_2$  &$0.267$     &$0.333$        &$0.242$      &$0.260$       &$0.349$ \\
\hline
PQCD\cite{PQCD}&$V$ &$0.318$     &$0.406$        &$0.305$      &              &        \\
\ \ \      & $A_0$  &$0.366$     &$0.455$        &$0.347$      &              &        \\
\ \ \      & $A_1$  &$0.25$      &$0.30$         &$0.24$       &              &        \\
\ \ \      & $A_2$  &$0.21$      &$0.24$         &$0.20$       &              &        \\
\hline
LQCD\cite{LQCD1}&$V$&$0.35 $     &               &             &              &        \\
\ \ \      & $A_0$  &$0.30 $     &               &             &              &        \\
\ \ \      & $A_1$  &$0.27$      &               &             &              &        \\
\ \ \      & $A_2$  &$0.26$      &               &             &              &        \\
\ \ \      & $T_1$  &         &$0.24$\cite{LQCD4}&             &              &        \\
\hline This work
\ \ \      & $V$    &$0.298$     &$0.339$        &$0.275$      &$0.323$       &$0.329$ \\
\ \ \      & $A_0$  &$0.260$     &$0.283$        &$0.240$      &$0.279$       &$0.279$ \\
\ \ \      & $A_1$  &$0.227$     &$0.248$        &$0.209$      &$0.228$       &$0.232$ \\
\ \ \      & $A_2$  &$0.215$     &$0.233$        &$0.198$      &$0.204$       &$0.210$ \\
\ \ \      & $T_1$  &$0.260$     &$0.290$        &$0.239$      &$0.271$       &$0.276$ \\
\ \ \      & $T_2$  &$0.260$     &$0.290$        &$0.239$      &$0.271$       &$0.276$ \\
\ \ \      & $T_3$  &$0.184$     &$0.194$        &$0.168$      &$0.165$       &$0.170$\\
 \hline \hline
\end{tabular}\label{physicalformPV}
\end{center}
\end{table}

The features of our results are:
\begin{itemize}

\item  Our results of $\zeta_{||}$ and $\zeta_\perp$ for every meson are close to
each other, which is mainly due to the similar wave function for the
longitudinal and transverse polarizations.

\item  The physical form factors can be directly calculated by
using the soft form factors. The kinematic factor as in Eq.~(\ref{reduction}) makes the
physical form factors different. $V$ is the largest form factor which is enhanced by the
factor $1+M_V/M_B$, while $T_3$ is smallest one for a minus term.

\item The soft form factors of $B\to K^*$  is larger than that of
$B\to\rho$ because  the $s$ quark  in $K^*$ meson carries more momentum than $d$ quark in
$\rho$,  which can induce more overlap of the $\bar B$ meson wave function and the light $K^*$
meson wave function. $\zeta_{||,\perp}^{B\to \omega}$ is smaller than
$\zeta_{||,\perp}^{B\to\rho}$, which is a consequence of the fact the decay constant of
$\omega$ is smaller than that of $\rho$.

\item As we have discussed above, we keep the first term in
$\zeta_{||}$, although it is suppressed by $\lambda= \sqrt{\Lambda_{QCD}/m_b}$. This term can
not be neglected in the numerics as the suppression is not so effective: the
$\zeta_{||}^{B\to\rho}$ without this term becomes: \be
\zeta_{||}^{B\to\rho}(\frac{m_B}{2})=0.139,\en which is quite smaller than the result with it.
This small $\zeta_{||}$  can lead to a small $A_0$ but a large $A_2$ and $T_3$.
\end{itemize}

\begin{figure}[[hbtp]
\vspace{-0.2cm}
\begin{center}
\includegraphics[scale=0.4]{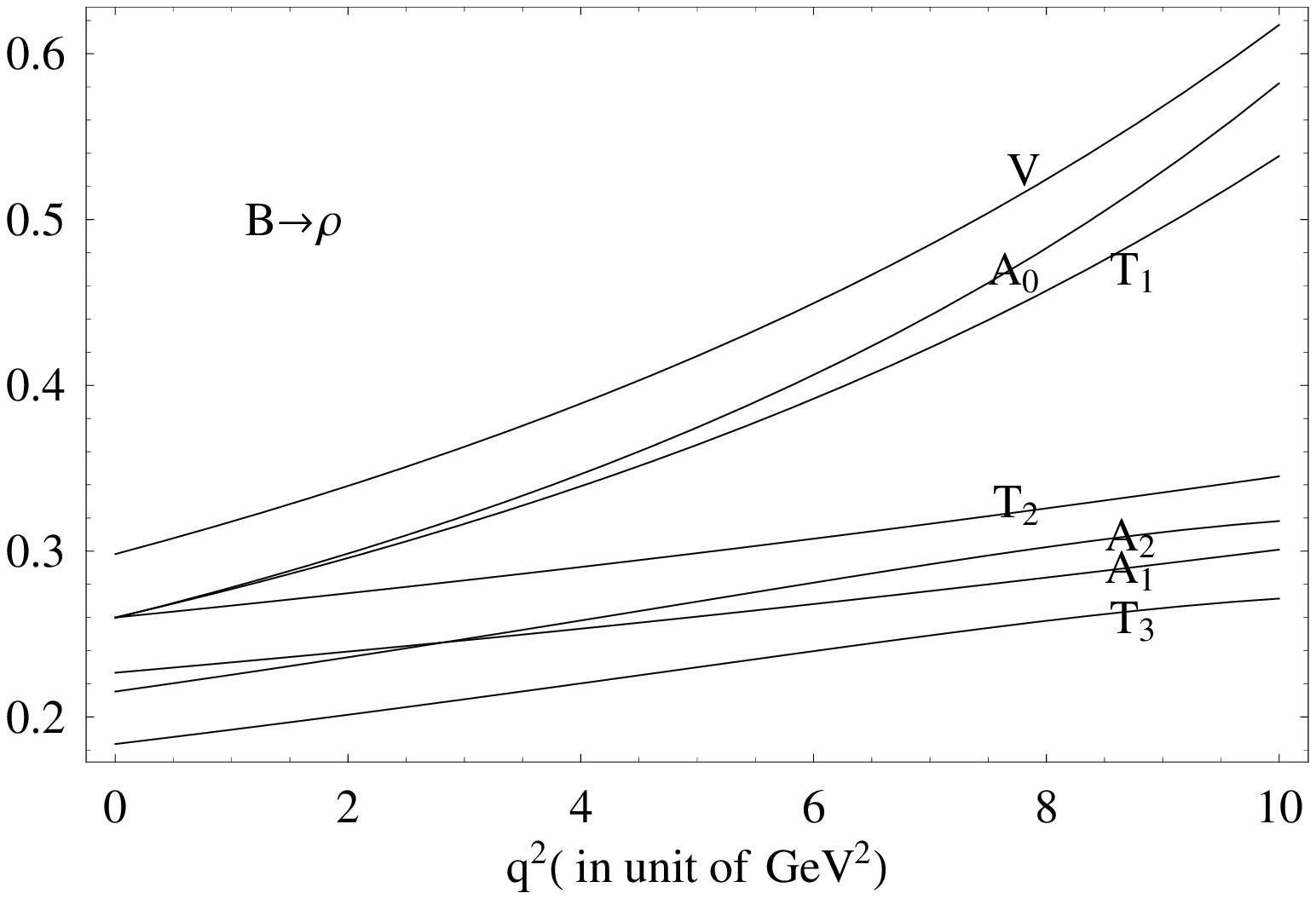}
\includegraphics[scale=0.4]{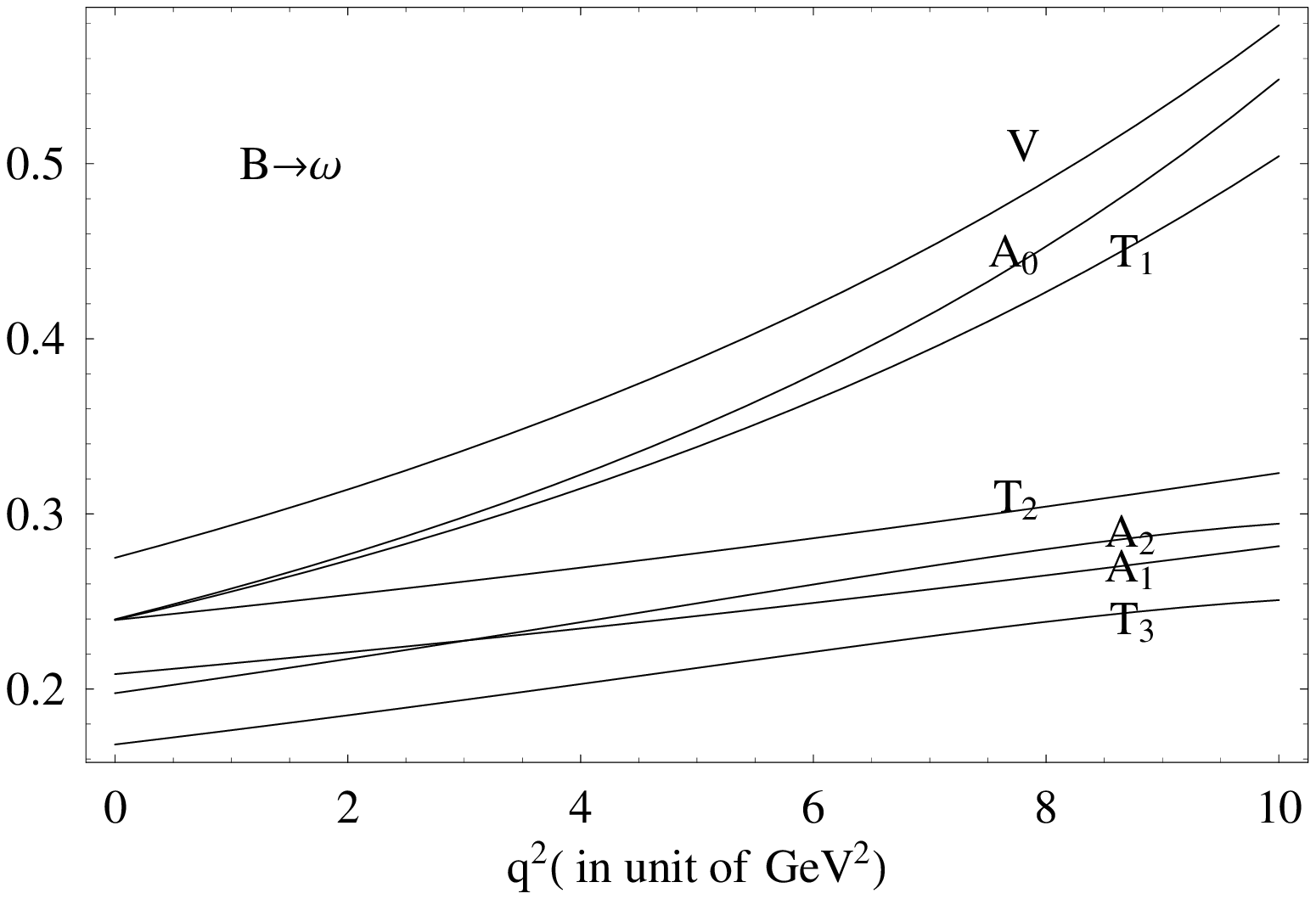}
\includegraphics[scale=0.42]{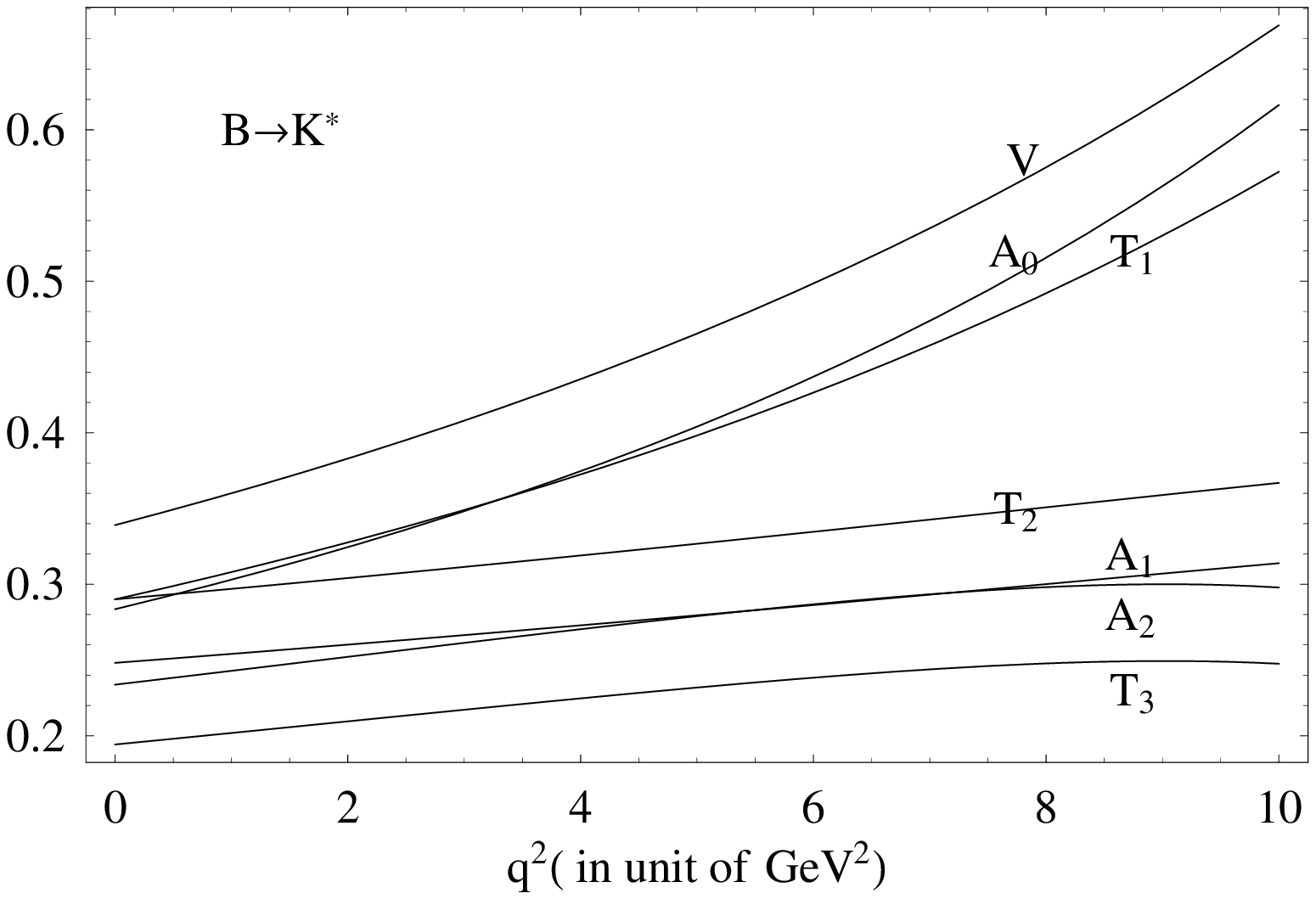}
\includegraphics[scale=0.4]{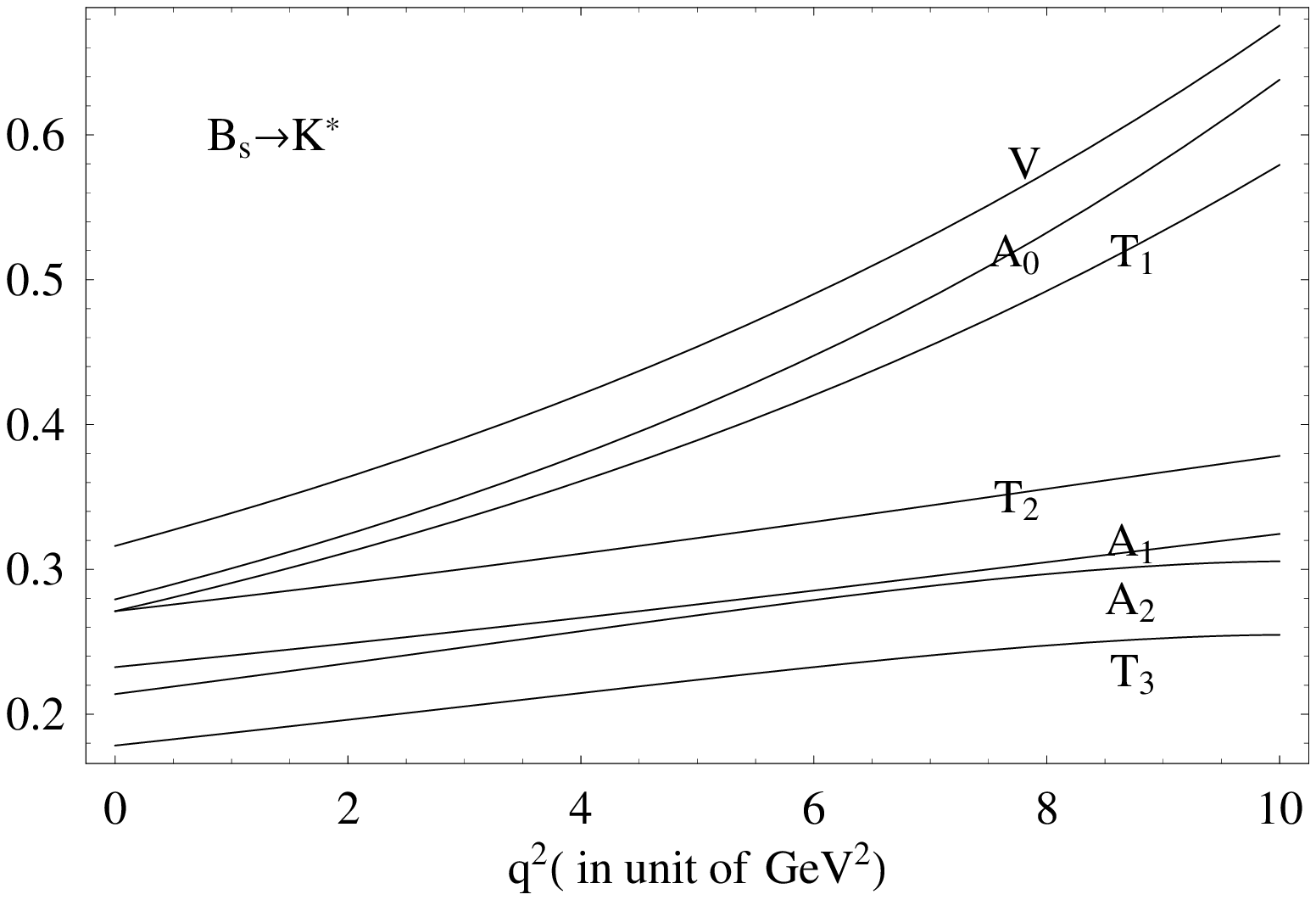}
\includegraphics[scale=0.4]{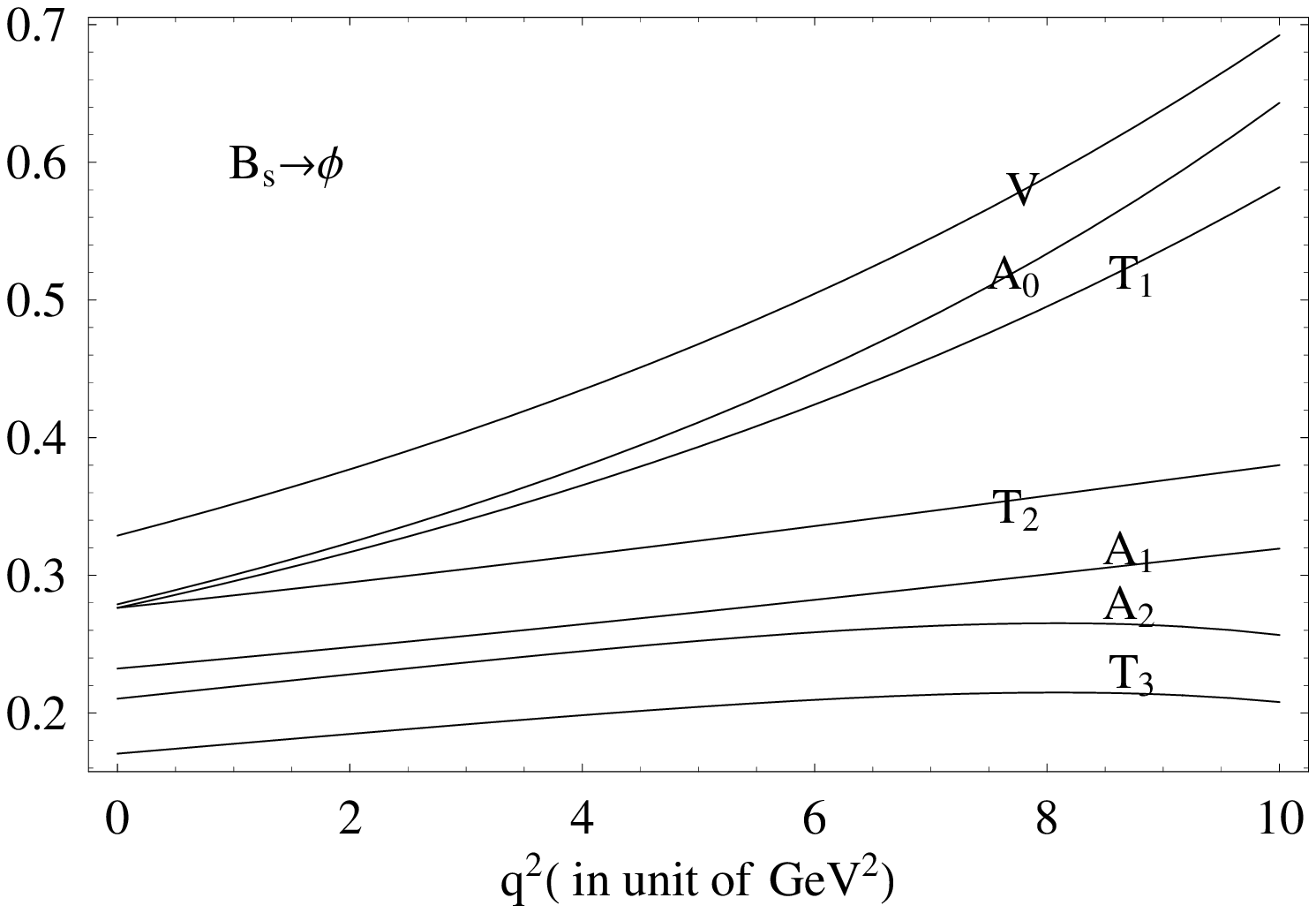}
\end{center}
\vspace{-0.2cm} \caption{{The $q^2$ dependence of the $B_{(s)}\to V$
form factors.}}\label{BVqdepend}
\end{figure}

The $q^2$ dependence ($0<q^2<10 \mbox{ GeV}^2$) of the form factors are plotted in
Fig.~\ref{BVqdepend}. The two form factors $V$ and $T_1$ have the same $q^2$ dependence except
the different results at $q^2=0$ and both of them can be directly calculated by
$\zeta_\perp(E)$. $A_0(q^2)=\zeta_{||}(E)$ has similar $q^2$ dependence with $\zeta_\perp(E)$.
When the $q^2$ gets large, $A_0$ is a little sharper than $V$ and $T_1$. The other four form
factors are rather flat and are less sensitive to $q^2$. From the figure, we can see that the
$A_2$ and $T_3$ show a tendency to decrease at large $q^2$, these two form factors may not be
described by the above parametrization and so we will not fit them as in $B$ to pseudoscalar
decays. We use the same parametrization to describe the $q^2$ dependence of the other form
factors, and the results for the fitted parameters are given in Table~\ref{zetaV}. From the
table, we can see that the parameters $a$ for various channels are close to each other: around
$a=1.5$ for $\zeta_{||}(A_0)$ and $\zeta_\perp(V,T_1)$ or $a=0.5$ for $\frac{2E}{m_B}
\zeta_\perp(A_1,T_2)$. Another interesting feature is that: for all form factors, the
parameter $b$ is not large and the form factor is dominated by the monopole term.

\begin{table}\caption{The parameters in the parametrization of
 $B\to V$ form factors.}
\begin{center}
\begin{tabular}{c c|c|c|c|c|c}
\hline \hline
 & & $\zeta_{||}^{B\to\rho}(A_0)$ & $\zeta_{||}^{B\to\omega}(A_0)$ & $\zeta_{||}^{B\to K^*}(A_0)$
   &$\zeta_{||}^{B_s\to K^*}(A_0)$   & $\zeta_{||}^{B_s\to\phi}(A_0)$ \\
\hline
  $a$       &            &$1.56$        &$ 1.60$        &$1.51$         &$1.74$         &$1.73$ \\
\hline
  $b$       &            & $0.17$       &$0.22$         & $0.14$        &$0.47$         &$0.41$ \\
\hline \hline
 & & $\zeta^{B\to\rho}_\perp(V,T_1)$&$\zeta^{B\to\omega}_\perp(V,T_1)$
   & $\zeta^{B\to K^*}_\perp(V,T_1)$
   &$\zeta^{B_s\to K^*}_\perp(V,T_1)$ & $\zeta^{B_s\to\phi}_\perp(V,T_1)$ \\
 \hline
  $a$       &            &$1.45$        &$ 1.49$        &$1.37$         &$1.64$         &$1.60$ \\
\hline
  $b$       &            & $0.15$       &$0.20$         &$0.11$         &$0.42$         &$0.36$ \\
\hline \hline
 & & $\frac{2E}{m_B}\zeta^{B\to\rho}_\perp(A_1,T_2)$
   & $\frac{2E}{m_B}\zeta^{B\to\omega}_\perp(A_1,T_2)$
   &$\frac{2E}{m_B}\zeta^{B\to K^*}_\perp(A_1,T_2)$
   & $\frac{2E}{m_B}\zeta^{B_s\to K^*}_\perp(A_1,T_2)$
   & $\frac{2E}{m_B}\zeta^{B_s\to\phi}_\perp(A_1,T_2)$ \\
 \hline
  $a$       &            &$0.62$        & $ 0.66$       &$0.55$         &$0.82$         &$0.48$ \\
\hline
  $b$       &            &$-0.11$       &$-0.10$        &$-0.05$        &$0.08$         &$0.04$ \\
\hline \hline
\end{tabular}\label{zetaV}
\end{center}
\end{table}

\section{Conclusions}

A light cone quark model within the soft collinear effective theory is constructed in this
study. We calculated all the heavy-to-light $B_{(s)}\to P$ and $B_{(s)}\to V$ transition form
factors at large recoil region. The three universal soft form factors are studied, in
particular, the $B\to V$ form factors $\zeta_{||,\bot}$ are given for the first time. Our
numerical results are in general consistent with other non-perturbative methods, such as light
cone sum rules and quark models within the theoretical errors.   The fact that our numerical
results are close to the results by other methods supports that the leading order soft
contribution is dominant in the light cone quark model. The theoretical uncertainties caused
by the less known $B$ meson decay constants are small. The $q^2$ dependence of the $B\to P,V$
form factor is also studied in the range $0<q^2<10\mbox{ GeV}^2$.

\section*{Acknowledgments}

We would like to thank F.K. Guo and F.Q. Wu for helpful discussions.
Z. Wei wishes to thank  the Institute of High Energy Physics for
their hospitality during his summer visit where part of this work
was done. This work was partly supported by the National Science
Foundation of China under Grant No.10475085 and 10625525.


\end{document}